\documentclass[10pt]{iopart}
\pdfoutput=1
\usepackage{graphicx}
\usepackage{xcolor}
\usepackage{iopams} 

\eqnobysec

\usepackage{color}
\def\CCB{\color{blue}}

\newcommand{\beq}{\begin{equation}}
\newcommand{\eeq}{\end{equation}}
\newcommand{\be}{\begin{equation*}}
\newcommand{\ee}{\end{equation*}}
\newcommand{\beqa}{\begin{eqnarray}}
\newcommand{\eeqa}{\end{eqnarray}}
\newcommand{\bea}{\begin{eqnarray*}}
\newcommand{\eea}{\end{eqnarray*}}

\newcommand{\comport}[3]{\mathrel{\mathop{#1}\limits_{#2}^{#3}}}
\def\binom#1#2{{#1\choose #2}}

\newcommand{\mean}[1]{\langle#1\rangle}
\newcommand{\frad}[2]{\displaystyle{\displaystyle#1\over\displaystyle#2}}

\newcommand{\prob}{\mathop{\rm Prob}\nolimits}
\newcommand{\var}{\mathop{\rm Var}\nolimits}

\begin{document}

\title{Condensation for random variables conditioned by the value of their sum}
\author{Claude Godr\`eche}
\address{Institut de Physique Th\'eorique, Universit\'e Paris-Saclay, CEA and CNRS, \\91191 Gif-sur-Yvette, France}

\begin{abstract}
We revisit the problem of condensation for independent, identically distributed random variables with a power-law tail, conditioned by the value of their sum.
For large values of the sum, and for a large number of summands, a condensation transition occurs where the largest summand accommodates the excess difference between the value of the sum and its mean.
This simple scenario of condensation underlies a number of studies in statistical physics, 
such as, e.g., in random allocation and urn models, random maps, zero-range processes and mass transport models.
Much of the effort here is devoted to presenting the subject in simple terms,
reproducing known results and adding some new ones.
In particular we address the question of the quantitative comparison between asymptotic estimates and exact finite-size results. 
Simply stated, one would like to know how accurate are the asymptotic estimates of the observables of interest, compared to their exact finite-size counterparts, to the extent that they are known.
This comparison, illustrated on the particular exemple of a distribution with L\'evy index equal to $3/2$, demonstrates the role of the contributions of the dip and large deviation regimes.
Except for the last section devoted to a brief review of extremal statistics, the presentation is self-contained and uses simple analytical methods.
\end{abstract}

\maketitle

\section{Introduction}

A question underlying a number of studies in statistical physics or in probability theory is the following.
Let $X_1,\dots, X_n$ be $n$ independent, identically distributed (iid) positive random variables 
with finite mean.
Assume that $n$ is large and that the sum of these random variables is conditioned to take a fixed value, which can be smaller, equal to or larger than its mean.
The question is to know how the (positive or negative) difference $\Delta$
between the fixed value of the sum and its mean is distributed amongst the summands $X_i$, once a dependency between them has been introduced by the conditioning.

The answer to this question can be informally summarised as follows.
If the common density of the random variables $X_i$ is exponential, then, after conditioning, each of the summands takes a bit of the difference $\Delta$, whether negative or positive.
The system is said to be in a `fluid phase'.
If this density is subexponential (power law, stretched exponential), the same holds when the difference $\Delta$ is negative.
However, 
when it is positive (i.e., in excess) and large, in contrast to the exponential case, in general \textit{only one} of the summands, the `condensate', bears this excess.
The remaining $n-1$ summands, which form the so-called `critical background', are essentially unconstrained.
This means that the dependency between the summands $X_i$ introduced by the conditioning goes asymptotically in the condensate.
One then speaks of a `condensation transition'.
When $\Delta=0$ the system is again essentially made of a critical background.

\begin{figure}[htb]
\begin{center}
\includegraphics[angle=0,width=1\linewidth]{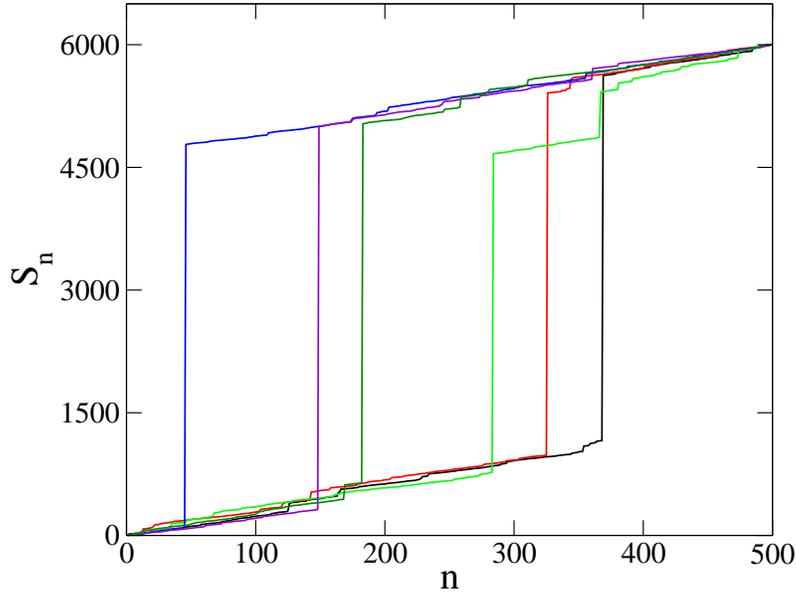}
\caption{\small
Six trajectories of a random walk $S_n=\sum_{i=1}^nX_i$ made of steps with power-law distribution $f_X(x)=3x^{-5/2}/2$ ($x>1$), for which $\mean{X}$=3.
The random walk is conditioned to end at position $4\mean{S_n}=6000$ at time $n=500$.
For each trajectory one can observe the occurrence of a `big jump' whose magnitude fluctuates around $\Delta=3\mean{S_n}=4500$.
Note however that the green history is made of two big jumps
(see section \ref{sec:marginal} for a discussion of this point).
}
\label{fig:RW}
\end{center}
\end{figure}

This phenomenon can be illustrated by considering a random walk whose steps are the summands $X_i$, and which is conditioned to end at a given position at time $n$.
Figure \ref{fig:RW} depicts six histories of such a random walk with a power-law distribution of steps with tail index $\theta=3/2$, conditioned to end at four times its mean, $4n\mean{X}$, at time $n$.
For each trajectory one can observe the occurrence of a `big jump' whose magnitude fluctuates around $\Delta=3n\mean{X}$.
In other words a large deviation of the sum is typically realised by a single big jump.
The latter, i.e., the greatest summand, is the condensate referred to above.
After removing this condensate the resulting histories are essentially unconstrained.
In figure \ref{fig:RW} one may note the presence of an history (in green) made of two big jumps.
The role of such trajectories will be 
discussed in section \ref{sec:marginal} and later sections.

The analytical formulation of this question is as follows.
The summands $X_i$ are, from now on, except at the end of this paper, continuous random variables.
Their common density is denoted by $f_X(x)$, with mean $\langle X\rangle\equiv c_{1}$ ($c_1$ is the first cumulant).
Denoting by $y$ the value taken by their sum, $S_n=\sum_{i}X_i$,
the joint density of the $X_i$ and of $S_n$ is 
\be
f(x_1,\dots,x_n,y)=f_X(x_1)\dots f_X(x_n)\delta\Big(\sum_{i=1}^nx_i-y\Big).
\ee
Summing upon all variables but $y$ yields
the density of $S_n$, 
\be
\fl f_{n}(y)=\frac{{\rm d}}{{\rm d}y}\prob(S_n<y)=\int {\rm d}x_1\dots{\rm d}x_n f_X(x_1)\dots f_X(x_n)
\delta\Big(\sum_{i=1}^nx_i-y\Big).
\ee
The joint conditional density of the random variables $X_1,\dots, X_n$ under the condition $S_n=y$, denoted for short by $f(x_1,\dots,x_n|y)$, therefore reads
\be
f(x_1,\dots,x_n|y)=\frac{f_X(x_1)\dots f_X(x_n)\delta(\sum_{i}x_i-y)}{f_{n}(y)},
\ee
the presence of the denominator ensuring the normalisation.

We shall mainly be interested in the marginal conditional distribution of one of the $X_i$, denoted for short by $f(x|y)$, obtained from the previous expression by summing upon all $X_i$ but one, to give
\beq\label{eq:marginal}
f(x|y)=f_X(x)\frac{f_{n-1}(y-x)}{f_{n}(y)},
\eeq
which can be interpreted as the ``dressed'' distribution of one of the $X_i$ as opposed to the `bare' distribution $f_X(x)$.
The associated conditional average is thus
\beq\label{eq:conditional}
\langle X|S_n=y\rangle=\int_0^{y}{\rm d}x\,f(x|y)=
\frac{y}{n}\equiv\rho.
\eeq
The difference $\Delta$ between the value of the sum $S_n$ and its mean $\mean{S_n}=nc_1$ 
can be therefore simply expressed in terms of the difference between the conditional and unconditional averages
\be
\Delta=y-nc_1=n(\rho-c_1)\equiv n[\langle X|S_n=y\rangle-\langle X\rangle].
\ee

Looking again at figure \ref {fig:RW}, the marginal $f(x|y)$ can be operationally seen as the limiting distribution of the summands (i.e., the step lengths of the random walk) for a large number of trajectories.
Since the largest summand, the condensate, appears to be clearly separated from the other ones, the marginal $f(x|y)$ is expected to have a hump shape in a neighbourhood of $\Delta$, representing the fluctuations of the condensate.

There are numerous studies related to this subject, dealing with urn models \cite{burda,burda2,camia,lux}, zero-range processes \cite{spitzer,andjel,evans1,jeon,cg,gross,evans2,gl2005,ferrari,gl2012,armendariz2009,armendariz2011,armendariz2013,lux}, mass-transport models \cite{maj1,maj2,maj3}, random allocation or random tree problems \cite{janson}, to quote but a few.
Large deviations for random walks with sub-exponential increments are considered in \cite{denisov}.

In the present work we revisit this very question with two more specific aims in mind.
Firstly, we shall devote special care to the analysis of the distribution of the sum, $f_n(y)$, and of the marginal distribution of the summands, $f(x|y)$, in the various regimes of interest, with emphasis on the role of rare events.
Secondly, for a particular example of power-law distribution of the summands, we shall confront the asymptotic predictions obtained for a large but finite number of summands to their exact counterparts.
This gives a hint of the accuracy of the predictions of asymptotic analysis for more general distributions where exact finite-size expressions are not available.

In what follows we focus on the case where the density $f_X(x)$ of the random variables $X_i$
has a power-law tail,
\beq\label{eq:powerlaw}
f_X(x)\comport{\approx}{x\to\infty}{} \frac{c}{x^{1+\theta}},
\eeq
with $\theta>1$ in order to have a finite mean $c_{1}$.
We shall however begin, in section \ref{sec:exponent}, by the analysis of the simpler situation where $f_X(x)$ is exponential, for which condensation does not occur.
We shall then proceed by analysing the general case of a power-law distribution (\ref{eq:powerlaw}).
As can be seen on the expression (\ref{eq:marginal}), the knowledge of the distribution of the sum, $f_n(y)$, allows to infer the marginal distribution $f(x|y)$.
The detailed analysis of $f_n(y)$ in the different regimes is therefore the building block for the study of the marginal $f(x|y)$ (section \ref{sec:power-law}).
This analysis will be done in Laplace space, using the preparatory material contained in section \ref{sec:laplace}.
The results thus obtained are then applied, in section \ref {sec:theta32},
 to the special instance of the distribution (\ref{eq:proto32}) with power-law exponent $\theta=3/2$, where exact expressions at finite $n$ can be derived, in order to illustrate and validate the asymptotic analysis made in the general case of section \ref{sec:power-law}.
Section \ref {sec:marginal} is devoted to the derivation of the marginal distribution $f(x|y)$ in the various regimes, both for a generic power-law distribution (\ref{eq:powerlaw}) and for the special case (\ref{eq:proto32}).
The question of the unicity of the condensate and the statistics of extremes are reviewed in sections \ref{sec:unique} and \ref{sec:extremes}.
The case of discrete random variables is summarised in \ref{app:discrete}.

The present study builds upon previous works, especially \cite{maj1,maj2,gl2005}, and
consists, to a large extent, of an update of \cite{maj2}, with some effort devoted to giving
a self-contained presentation, using simple analytical methods.
It has no pretension to being exhaustive on all aspects of the field.
In particular, reviewing the vast mathematical literature on sums of iid subexponential random variables and on the distribution of such random variables conditioned by a large value of their sum is beyond the scope of this work.
The mathematical references most relevant to the present work are \cite{jeon,ferrari,armendariz2009,armendariz2011,armendariz2013,janson}, mentioned above.
Let us finally mention \cite{chleboun}, devoted to finite-size effects in zero-range condensation as manifested for example in the current overshoot, which shares some common features with the present work.
\section{Exponentially distributed iid random variables}
\label{sec:exponent}

We start with the simple case of the exponential distribution
\be
f_X(x)=\frac{\e^{-x/ c_{1}}}{ c_{1}},
\ee
for which the distribution of the sum $f_n(y)$ and the marginal $f(x|y)$ are known exactly.
First, the sum, $S_n$, has a gamma distribution
\beq\label{eq:gamma-}
f_{n}(y)=\frac{y^{n-1}\e^{-y/ c_{1}}}{c_{1}^n\,\Gamma(n) },
\eeq
which is the inverse Laplace transform (with $\mathrm{Re}\ s>-1/c_1$) of
\be
\hat f_{n}(s)=(\hat f_X(s))^n=\frac{1}{(1+s c_{1})^n},
\ee
as can be checked by inspection.
Therefore the marginal distribution $f(x|y)$ (\ref{eq:marginal}) is inferred from the exact expression
 (\ref{eq:gamma-}) to give
\beq\label{margin-exact-}
f(x|y)=(n-1)\frac{(y-x)^{n-2}}{y^{n-1}}.
\eeq
It does not depend on $c_1$ and is monotonically decreasing with $x$, which is a manifestation of the absence of condensation.
The conditional average $\langle X|S_n=y\rangle$ (\ref{eq:conditional}) computed from (\ref{margin-exact-}) is equal to $\rho$, as it should.

Setting $y=n\rho$ in (\ref{margin-exact-}) and letting $n\to\infty$, with $\rho$ and $x$ fixed yields the asymptotic estimate\footnote{The symbol $\approx$ stands for asymptotic equivalence.
The symbol $\sim$ stands either for `of the order of', or for
`with exponential accuracy'.}
\beq\label{margin-exp-}
f(x|y)\approx \frac{n}{y}(1-x/y)^{n}
\approx\frac{\e^{-x/\rho}}{\rho}.
\eeq
This estimate holds irrespectively of whether $\rho$ is smaller or larger than $\mean{X}=c_{1}$.
In other words, the system adjusts itself in such a way that the conditional distribution $f(x|y)$ is still given by the `bare' distribution, $f_X(x)$, with only a change of the parameter from $c_1$ to $\rho\lessgtr c_1$.

We now turn to the large deviation estimate of $f_n(y)$.
We set, as above, $y=n\rho$ in the expression (\ref{eq:gamma-}) of $f_n(y)$ and take the limit $n\to\infty$.
This yields
\beq\label{eq:gde-dev-exp2-}
f_{n}(y)\approx \frac{\e^{n(1-\rho/c_1+\ln \rho/c_1)}}{\sqrt{2\pi n}\,\rho},
\eeq
which reproduces the exact distribution (\ref{eq:gamma-}) up to the replacement of $\Gamma(n)$ by its Stirling approximation.
With exponential accuracy we can write
\be
f_n(y)\sim\e^{-nI(\rho)},
\ee
where the large deviation function,
\beq\label{LDF-exp}
-I(\rho)=1-\frac{\rho}{ c_{1}}+\ln \frac{\rho}{ c_{1}},
\eeq
is defined for any value of the density $\rho$ and is minimal and vanishes at $\rho= c_{1}$.

Using (\ref{LDF-exp}) yields an accurate estimate of $f(x|y)$ for all values of $x$.
In particular, (\ref{margin-exp-}) is recovered in the same limit as above,
setting $y=n\rho$ and letting $n\to\infty$, for $\rho$ and $x$ fixed.

Anticipating on the sequel (compare to (\ref{eq:marginal-sub})), the rightmost expression in (\ref{margin-exp-}) can be recast as
\beq\label{eq:marginal-sub0}
f(x|y)\approx\frac{ \e^{- s_{\rho} x}f_X(x)}{\int_0^\infty {\rm d}x\, \e^{- s_{\rho} x}f_X(x)},
\eeq
where
\be
 s_{\rho}=\frac{1}{\rho}-\frac{1}{c_1}
\ee
can be positive, negative or zero.
The denominator in (\ref{eq:marginal-sub0}) ensures normalisation.
If we use (\ref{eq:marginal-sub0}) to compute the density $\rho$ by (\ref{eq:conditional}), we find
a relation between $ s_{\rho}$ and $\rho$,
\beq\label{eq:magic}
\rho\approx\frac{\mean{X \e^{- s_{\rho} X}}}{\mean{ \e^{- s_{\rho} X}}}.
\eeq
As shown later, (\ref{eq:magic}) is the saddle-point equation for the inverse Laplace representation of $f_n(y)$.

To conclude, there is no condensation in the present case.
The system is always in a fluid phase where, irrespectively of its sign, the difference $\Delta$
is evenly distributed over all summands.

\section{Laplace space and singularities}
\label{sec:laplace}
In what follows the asymptotic analysis of the distribution $f_n(y)$ of the sum $S_n$ is performed in Laplace space.
The Laplace transform of $f_n(y)$ with respect to $y$ is
\be
\hat f_{n}(s)=(\hat f_X(s))^n,
\ee
where $\hat f_X(s)\equiv\langle \e^{-sX}\rangle$,
hence, by inversion,
\beq\label{eq:fSny}
f_{n}(y)=\int_{C}\frac{{\rm d}s}{2\pi{\rm i}}\e^{sy}(\hat f_X(s))^n,
\eeq
where $C$ is a Bromwich contour located on the right of the origin.
The analysis of the distribution of the sum $S_n$ therefore relies upon the analysis of the singularities of $\hat f_X(s)$ in the complex $s-$plane.
For the power-law distribution (\ref{eq:powerlaw}) the Laplace transform $\hat f_X(s)$ has a cut extending along the negative real axis.
When $n$ is large $(\hat f_{X}(s))^n$ is dominated by $\hat f_X(s)\approx 1$, i.e., $s$ small.
The analytical structure of the Laplace transform $\hat f_X(s)$ in the vicinity of the origin will therefore play a crucial role in the analysis of the distribution of $S_n$.

For a density $f_X(x)$ with a power-law tail (\ref{eq:powerlaw}) the expansion of $\hat f_X(s)$ for $s\to0$, can be decomposed into a regular and a singular part
\beqa\label{eq:regul}
\hat f_{\rm reg}(s)=1-s c_{1}+\frac{s^2}{2}\langle X^2\rangle+\cdots,
\\
\hat f_{\rm sing}(s)=a s^{\theta}+\cdots,
\label{eq:singul}
\eeqa
where the parameter $a$ is related to the tail parameter $c$ by \cite{zolo}
\beq\label{eq:a-c}
a=\Gamma(-\theta)\,c.
\eeq
The parameter $a$ is negative if $0<\theta<1$, positive if $1<\theta<2$, and so on.
For instance, $\Gamma(-1/2)=-2\sqrt{\pi}$, $\Gamma(-3/2)=4\sqrt{\pi}/3$, $\Gamma(-5/2)=-8\sqrt{\pi}/15$.
The number of non-zero moments in the expansion of the regular part depends on the value of $\theta$.
For $1<\theta<2$ the first moment is defined, for $2<\theta<3$ the second moment is also defined, and so on.
The expansion of the generating function of cumulants $K(s)=\ln \hat f_X(s)$ 
follows from (\ref{eq:regul}) and (\ref{eq:singul}) 
\beq\label{eq:kexpans}
K(s)=-s c_{1}+\frac{s^2}{2}c_2-\cdots+as^{\theta}+\cdots,
\eeq
where 
\be
c_2\equiv\var X,
\ee
denotes the second cumulant.
The first dots stand for higher-order regular terms
($ s^3$,~$\dots$)
and the second dots stand for higher-order singular terms ($s^{\theta+1}$, $\dots$).

\section{Sum of iid positive random variables with a power-law tail}
\label{sec:power-law}

We now focus on the case where the density $f_X(x)$ has a power-law tail (\ref{eq:powerlaw}) with exponent $\theta$.
We will investigate successively the bulk of the distribution of $S_n$ (generalised central limit theorem), then its left and right tails.

\subsection{Generalised central limit theorem}

\subsubsection*{\CCB Reminder.}
We start with a reminder of well-known results on the generalised central limit theorem.
By completeness we consider also the case where $\theta<1$, though it is not relevant for the present study since the first moment $\mean{X}=c_1$ is infinite.

The generalised central limit theorem \cite{gnedenko} states that, for iid random variables with density (\ref{eq:powerlaw}), there exists two positive sequences $a_n$ and $b_n$ such that, when $n\to\infty$,
the centered and scaled sum 
\be
U_n= \frac{S_n-b_n}{a_n}
\ee
converges (in distribution) to a stable law with index $\alpha$, where
\beq\label{eq:alfa}
\alpha= \left\{
\begin{array}{ll}
\theta & \textrm{ if } \theta<2, \vspace{2pt}
\\ 
2 & \textrm{ if } \theta>2,
\end{array}
\right.
\eeq
and asymmetry parameter $\beta=1$.
Indeed, in the general case of a distribution $f_X(x)$ with right and left power-law tails $c_{\pm}/|x|^{1+\theta}$ ($x\to\pm\infty$), the asymmetry parameter $\beta$ is, by definition, the ratio $(c_{+}-c_{-})/(c_{+}+c_{-})$.
In the present case of positive random variables the parameter $c_{-}=0$, and $\beta$ is thus equal to unity.
We denote $c_{+}$ by $c$, as in (\ref{eq:powerlaw}).
If $0<\alpha<2$, this stable law also depends on the tail parameter $c$.
If $\alpha=2$ the stable law is a Gaussian, the expression of which neither contains the asymmetry parameter $\beta$ nor the tail parameter $c$.

The scale parameter $a_n$ is equal to $n^{1/\alpha}$, where $\alpha$ is given by (\ref{eq:alfa}), the centering parameter $b_n$ is equal to $nc_1$ when the mean is finite ($\theta>1$), and to zero otherwise ($0<\theta<1$).
Thus, for $\theta>2$ ($\alpha=2$), the usual central limit theorem is recovered,
\beq\label{eq:Gauss}
\prob(u_1\le U_n\le u_2)\comport{\longrightarrow}{n\to\infty}{}\frac{1}{\sqrt{2\pi c_2}}\int_{u_1}^{u_2}{\rm d}u\, \e^{-u^2/2 c_2},
\eeq
while for $0<\theta<2$ ($\alpha=\theta$), the generalised central limit theorem reads
\be
\prob(u_1\le U_n\le u_2)\comport{\longrightarrow}{n\to\infty}{}\int_{u_1}^{u_2}{\rm d}u\, L_{\alpha,c}(u),
\ee
where $L_{\alpha,c}(u)$ is the density of the stable law of index $\alpha$, asymmetry parameter $\beta=1$ and tail parameter $c$.
To summarise, the (generalised) central limit theorem gives the universal behaviour of the distribution of the sum $S_n$ in the bulk, namely
\beq\label{eq:gclt1}
f_{n}(y)\approx\frac{1}{n^{1/2}}G\left(\frac{y-n c_{1}}{n^{1/2}}\right),
\eeq
if $\theta>2$ $ (\alpha=2)$,
where $G(u)$ is the Gaussian defined in (\ref{eq:Gauss}),
\beq\label{eq:gclt2}
f_{n}(y)\approx\frac{1}{n^{1/\theta}}L_{\theta,c}\left(\frac{y-n c_{1}}{n^{1/\theta}}\right),
\eeq
if $1<\theta<2$ $(\alpha=\theta)$, and
\beq\label{eq:01}
f_{n}(y)\approx\frac{1}{n^{1/\theta}}L_{\theta,c}\left(\frac{y}{n^{1/\theta}}\right),
\eeq
if $0<\theta<1$ $ (\alpha=\theta)$.

\subsubsection*{\CCB Examples.}
For instance, for $\alpha=1/2$, this distribution, the so-called L\'evy law of index $1/2$, is explicit and reads
\beqa\label{eq:Levy}
L_{1/2,c}(u)=\frac{c\,\e^{-\pi c^2/u}}{u^{3/2}},\quad (u>0),
\\
\hat L_{1/2,c}(s)=\e^{-2c\sqrt{\pi s}}.\label{eq:Levys}
\eeqa
Another example, analysed in detail later, is the stable law with index $\alpha=3/2$, which is explicitly given in terms of the Airy function (see (\ref{eq:airy})).
More generally the Laplace transform of any stable law with index $0<\alpha<2$ ($\alpha\ne1$) and asymmetry parameter $\beta=1$ reads
\beq\label{eq:stable}
\hat L_{\alpha,c}(s)=\e^{a s^\alpha},
\eeq
where the parameter $a$ is defined in (\ref{eq:a-c}).
Thus in direct space
\beq\label{eq:levy-alph}
L_{\alpha,c}(u)=\int_{C}\frac{{\rm d}s}{2\pi{\rm i}}\e^{s u+a s^{\alpha}},
\eeq
where $C$ is a Bromwich contour located on the right of the origin.
For $0<\alpha<1$ the density of the stable law is only defined for $u>0$, while for $1<\alpha<2$ the support of the density is the whole real axis, implying that its Laplace transform is bilateral. 

\subsubsection*{\CCB Short proof of the generalised central limit theorem.}

We start with the case $1<\theta<2$.
The generating function of cumulants $K(s)$ is, for small $s$, keeping the leading terms,
\beq\label{eq:ks-expansion}
K(s)=\ln \hat f_X(s)\approx-s c_{1}+as^{\theta},
\eeq
so, in this regime, the estimate of (\ref{eq:fSny}) is
\beq\label{eq:lapinverse}
f_{n}(y)\approx\int_{C}\frac{{\rm d}s}{2\pi{\rm i}}\e^{s(y-n c_{1})+n a s^{\theta}}.
\eeq
Setting 
\beq\label{eq:regime}
 y-nc_1=u\,n^{1/\theta},\qquad s=t\,n^{-1/\theta},
\eeq
yields (\ref{eq:gclt2}), using (\ref{eq:levy-alph}).
The regime considered here thus corresponds to 
$\rho\to c_{1}$.
We proceed likewise for $\theta>2$.
Keeping the leading terms in the expansion of $K(s)$, we obtain
\beq\label{eq:lapinverse+}
f_{n}(y)\approx\int_{C}\frac{{\rm d}s}{2\pi{\rm i}}\e^{s(y-n c_{1})+n c_2 s^2/2}.
\eeq
We now set
\beq\label{eq:regimeTLC}
y-nc_1=u\sqrt{n},\qquad s=t/\sqrt{n},
\eeq
which leads to the usual central limit theorem (\ref{eq:gclt1}).
The third case (\ref{eq:01}) can be proven likewise.

\subsubsection*{\CCB Asymptotic behaviours of stable laws.}
In both cases (i.e., if either $0<\alpha<1$ or $1<\alpha<2$) $L_{\alpha,c}(u)$ has the same right tail (\ref{eq:powerlaw}) as the initial distribution $f_X(x)$,
\beq\label{eq:tail-alph}
L_{\alpha,c}(u)\comport{\approx}{u\to\infty}{}\frac{c}{u^{1+\alpha}}, \quad (0<\alpha<2),
\eeq
as can be seen by linearising the integrand of (\ref{eq:levy-alph}) with respect to $s^{\alpha }$, and folding the contour around the negative real axis
(see for details in section \ref{sec:deepin} where the same reasoning is used).

The asymptotic behaviour of the stable law on the left can be obtained by the saddle-point method.
We have
\beq\label{eq:asympu0}
L_{\alpha,c}(u)\comport{\approx}{u\to0}{}\frac{B}{\,u^{\nu}}\e^{-A /u^{\mu}},
\quad (0<\alpha<1),
\eeq
\beq\label{eq:asymp-neg}
L_{\alpha,c}(u)\comport{\approx}{u\to-\infty}{}B |u|^{\nu}\e^{-A|u|^{\mu}},
\quad (1<\alpha<2),
\eeq
with exponents
\be
\mu=\frac{\alpha}{|1-\alpha|},\quad\nu=\frac{2-\alpha}{2|1-\alpha|},
\ee
and where the two positive constants $A$ and $B$ read
\be
A=\frac{|1-\alpha|}{\alpha}\left(\alpha|a|\right)^{1/(1-\alpha)},\quad 
B=\frac{(\alpha|a|)^{1/(2(1-\alpha))}}{\sqrt{2\pi |1-\alpha|}}.
\ee
For example, if $\alpha=1/2$, the asymptotic estimate (\ref{eq:asympu0}) reproduces identically the whole law (\ref{eq:Levy}).
For $\alpha=3/2$ we obtain, using (\ref{eq:a-c}),
\beq\label{eq:Lev32as}
L_{3/2,c}(u)\comport{\approx}{u\to-\infty}{}\frac{\sqrt{|u|}}{2\pi c}\e^{-|u|^3/(12\pi c^2)},
\eeq
a result related to (\ref{eq:gclt32}) below.

\subsubsection*{\CCB Away from the bulk.}
The generalised central limit theorem does not predict the behaviour of the distribution of the sum $S_n$ in the tails. 
We now investigate the behaviour of $f_{n}(y)$ away from the bulk, that is, when the difference $|\Delta|=|y-n c_{1}|$ is extensive, i.e., of order $n$, (while in the regimes (\ref{eq:regime} and (\ref{eq:regimeTLC}) it was subextensive), first to the left ($y<n c_{1}$), then to the right ($y>n c_{1}$), restricting the study to the case $\theta>1$, such that $ c_{1}$ is finite.

\subsection{Left tail: large deviations}
\label{sec:leftLD}

The left tail of $f_{n}(y)$ corresponds to those rare events where $\rho< c_{1}$, hence $\Delta$ large and negative.
In this regime, which is far away from the regime of validity of the generalised central limit theorem, the large deviation estimate of the density $f_{n}(y)$ is non universal and depends on the details of the distribution $f_X(x)$.
We first present the general framework for the computation of the large deviation function $I(\rho)$ (\ref{eq:largedev01}), valid for any $\theta>1$.
There is no explicit expression of this function in general for distribution of the type (\ref{eq:powerlaw}).
We shall later find an explicit expression of this large deviation function for the distribution (\ref{eq:proto32}) with tail index $\theta=3/2$, valid in all regimes (see section \ref{sec:theta32}).
For the time being, we will content ourselves with the expressions (\ref{eq:largedev+}) and (\ref{eq:CLTback}) of the large deviation function in the scaling regime where $\rho$ is close to $ c_{1}$, for a general distribution (\ref{eq:powerlaw}).
Equation (\ref{eq:largedev+}) restores the generalised central limit theorem in the regime (\ref{eq:asymp-neg}).
Equation (\ref{eq:CLTback}) restores the usual central limit theorem.

\subsubsection*{\CCB General framework.} 

Let us come back on (\ref{eq:fSny}) that we recast as
\beq\label{eq:fSny+}
f_{n}(y)=\int_{C}\frac{{\rm d}s}{2\pi{\rm i}}\e^{sy+n K(s)}=
\int_{C}\frac{{\rm d}s}{2\pi{\rm i}}\e^{-nI(\rho,s)},
\eeq
with
\beq\label{eq:Irhos}
-I(\rho,s)=s\rho +K(s).
\eeq
If $n$ is large it is natural to perform a saddle-point analysis of (\ref{eq:fSny+}).
The saddle-point equation reads
\be
\left. \frac{{\rm d}I(\rho,s)}{{\rm d}s}\right |_{ s_{\rho}}=0,
\ee
that is to say\footnote{The saddle-point equation (\ref{eq:saddleGen}) was anticipated in (\ref{eq:magic}).}
\beq\label{eq:saddleGen}
-K'( s_{\rho})
=\frac{\mean{X\e^{- s_{\rho} X}}}{\mean{\e^{- s_{\rho} X}}}
=\rho.
\eeq
The position of the saddle point $ s_{\rho}$ on the real axis depends on the value of $\rho=y/n$. 
This saddle point only exists if $\rho< c_{1}$.
Indeed, if $\rho= c_{1}$, 
the saddle point $ s_{\rho}=0$ hits the head of the cut of $K(s)$ (see (\ref{eq:kexpans})),
hence the saddle-point equation (\ref{eq:saddleGen}) cannot be satisfied beyond $\rho=c_1$.
Defining the large deviation function as
\beq\label{eq:largedev01}
-I(\rho)\equiv-I(\rho, s_{\rho})= s_{\rho}\rho+K( s_{\rho}),
\eeq
we finally obtain
\beq\label{eq:largedev02}
f_{n}(y)\comport{\approx}{n\to\infty}{} \frac{\e^{-nI(\rho)}}{\sqrt{2\pi nK''( s_{\rho})}}, \qquad (\rho< c_{1}),
\eeq
with
\beq\label{eq:Kpp}
K''(s)=\frac{\mean{X^2\e^{-s X}}}{\mean{\e^{-s X}}}-\Big(\frac{\mean{X\e^{-s X}}}{\mean{\e^{-s X}}}\Big)^2.
\eeq

\subsubsection*{\CCB Scaling regime.}
\label{sec:scalreg}
Determining the large deviation function in the scaling region $\rho\to c_{1}$
implies expanding the expressions above for $ s_{\rho}\to0$.
\\
$\bullet$ We start with $1<\theta<2$.
The saddle-point equation
\be
-K'( s_{\rho})\approx c_{1}-a\theta  s_{\rho}^{\theta-1}=\rho,
\ee
yields
\beq\label{eq:sstar}
 s_{\rho}\approx \left(\frac{ c_{1}-\rho}{a\theta}\right)^{1/(\theta-1)},
\eeq
only defined if $\rho< c_{1}$.
We thus find, using (\ref{eq:ks-expansion}), the expression of the large deviation function in this regime,
\beqa\label{eq:largedev+}
I(\rho)
\approx
\frac{\theta-1}{\theta(a\theta)^{1/(\theta-1)}}( c_{1}-\rho)^{\theta/(\theta-1)}.
\eeqa
The right side of this equation can be identified with $A|u|^\mu/n$ in (\ref{eq:asymp-neg}).
Actually, in this scaling regime, the full large deviation estimate (\ref{eq:largedev02}) reduces to (\ref{eq:gclt2}) with (\ref{eq:asymp-neg}).
The left tail (\ref{eq:asymp-neg}) can indeed be seen as the large deviation estimate of $L_{\theta,c}(u)$.

The special case of $\theta=3/2$ is treated in detail in section \ref{sec:theta32}.
The universal part of the large deviation function (\ref{eq:largedev+}) gives (\ref{eq:Irhobulk}).
\\
$\bullet$ For $\theta>2$ we have
$-K'( s_{\rho})\approx c_{1}- c_2  s_{\rho}=\rho$,
hence $ s_{\rho}\approx( c_{1}-\rho)/ c_2$,
yielding the quadratic form 
\beq\label{eq:CLTback}
I(\rho)\approx \frac{( c_{1}-\rho)^2}{2 c_2}.
\eeq
Thus (\ref{eq:largedev02}),
with $K''( s_{\rho})\approx c_2$, gives the central limit theorem (\ref{eq:gclt1}) back.

\subsubsection*{\CCB Remark.}
The two equations (\ref{eq:largedev01}) and (\ref{eq:largedev02}) provide a parametric representation of $I(\rho)$, i.e., of $K'( s_{\rho}) s_{\rho}-K( s_{\rho})$ against $-K'( s_{\rho})$ (which is $\rho$),
which can be used for numerical purposes.

\subsection{Right tail: `deep in the condensed phase'}
\label{sec:deepin}

Again the regime considered here, where $\rho> c_{1}$, is different from that prevailing for the central limit theorem. 
Recall that, for any value of $\theta>1$, using (\ref{eq:kexpans}),
\beq\label{eq:deepin}
f_{n}(y)=\int_{\rm C}\frac{{\rm d}s}{2\pi{\rm i}}
\e^{s(y-n c_{1})+\cdots+n a s^{\theta}+\cdots}.
\eeq
Now $\Delta=n(\rho-c_1)$ is of order $n$, so $s\sim 1/n$, implying that $ns^\theta\sim n^{1-\theta}$ is subextensive.
Therefore the two terms $s(y-nc_1)$ and $nas^\theta$ are no longer balanced as in (\ref{eq:lapinverse}) and (\ref{eq:regime}).
The contour ${\rm C}$ is deformed to encircle the real negative axis.
The leading contribution to $f_{n}(y)$ comes from linearising
with respect to the leading singular term:
\be
f_{n}(y)\approx na\int_{\rm C}\frac{{\rm d}s}{2\pi{\rm i}}\,\e^{s(y-n c_{1})}\, s^{\theta}.
\ee
Using the Hankel representation of the reciprocal Gamma function
\be
\frac{1}{\Gamma(\theta)}=\int_{\rm C}\frac{{\rm d}s}{2\pi{\rm i}}\e^s s^{-\theta},
\ee
we obtain
\be
f_{n}(y)\approx \frac{n a}{\Gamma(-\theta)\,(y-n c_{1})^{1+\theta}}.
\ee
Finally, using (\ref{eq:a-c}), we have, for any value of $\theta>1$, if $\rho> c_{1}$, thence for $y-n c_{1}\sim n$,
\beq\label{eq:deep}
f_{n}(y)\approx
\frac{nc}{(y-n c_{1})^{1+\theta}},
\eeq
where $c$ is the tail coefficient of $f_X(x)$.
Similar considerations can be found in \cite{gl2005,evans2}.

This result matches with the asymptotic estimate (\ref{eq:tail-alph}) for $y-n c_{1}= n^{1/\theta} u$ ($u$ large), if $1<\theta<2$ (see (\ref{eq:gclt2})).
This prediction holds further away in the tail, where the excess difference is extensive.
Furthermore (\ref{eq:deep}) also holds for $\theta>2$.
In other words, while at the scale $y-n c_{1}\sim n^{1/2}$ the tail is Gaussian, at the scale $y-n c_{1}\sim n$ it is given by (\ref{eq:deep}).
Equating (\ref{eq:gclt1}) and (\ref{eq:deep}) shows that the matching between the two behaviours occurs for
\beq\label{eq:matching}
 y-n c_{1}\sim \sqrt{(\theta-2)c_2}\,\sqrt{n\ln n}.
\eeq
See \cite{armendariz2013} for related considerations.

As a last comment, let us remark that the contributions coming from the next terms $(nas^\theta)^p$ in the expansion of $\e^{nas^\theta}$ in (\ref{eq:deepin}) are subleading by successive factors $n^{-(p-1)(\theta-1)}$ with respect to the contribution of the first term $nas^\theta$.
These subleading probabilities will be recovered otherwise in section \ref{sec:unique}.

\section{The example of a distribution with power-law tail exponent $\theta=3/2$}
\label{sec:theta32}

For the distribution
\beq\label{eq:proto32}
f_X(x)=\frac{2}{\sqrt{\pi}}\frac{\e^{-1/x}}{x^{5/2}},\qquad (x\ge0),
\eeq
such that $\langle X\rangle\equiv c_{1}=2$,
the exact distribution of the sum $S_n$ is explicit and reads \cite{maj2}
\beq\label{eq:hermit}
f_{n}(y)=\frac{n\,\e^{-n^2/y}}{\sqrt{\pi}y^{(n+3)/2}}
\left[H_n\left(\frac{y+2n}{2\sqrt{y}}\right)-\sqrt{y}H_{n-1}\left(\frac{y+2n}{2\sqrt{y}}\right)\right],
\eeq
where the $H_n$ are Hermite polynomials.
This exact result will provide an illustration of the statements made in the previous section as well as a benchmark for the asymptotic estimates given there.
In Laplace space
\be
\hat f_X(s)=(1+2\sqrt{s})\e^{-2\sqrt{s}},
\ee
as can be found by taking the derivative of (\ref{eq:Levy}) and (\ref{eq:Levys}) with respect to the tail parameter $c$.
So, for small $s$,
\beq\label{eq:fXs32}
\hat f_X(s)\approx1-2s+\frac{8}{3}s^{3/2},
\eeq
which is the beginning of the expansion $1-s c_{1}+a s^{\theta}+\cdots$, with $ c_{1}=2$ and $a=8/3$ obtained from (\ref{eq:a-c}) for $c=2/\sqrt{\pi}$ and $\theta=3/2$.
The generating function of cumulants is thus equal to
\be
K(s)=\ln \hat f_X(s)\approx-2s+\frac{8}{3}s^{3/2}.
\ee

\begin{figure}[htb]
\begin{center}
\includegraphics[angle=0,width=1\linewidth]{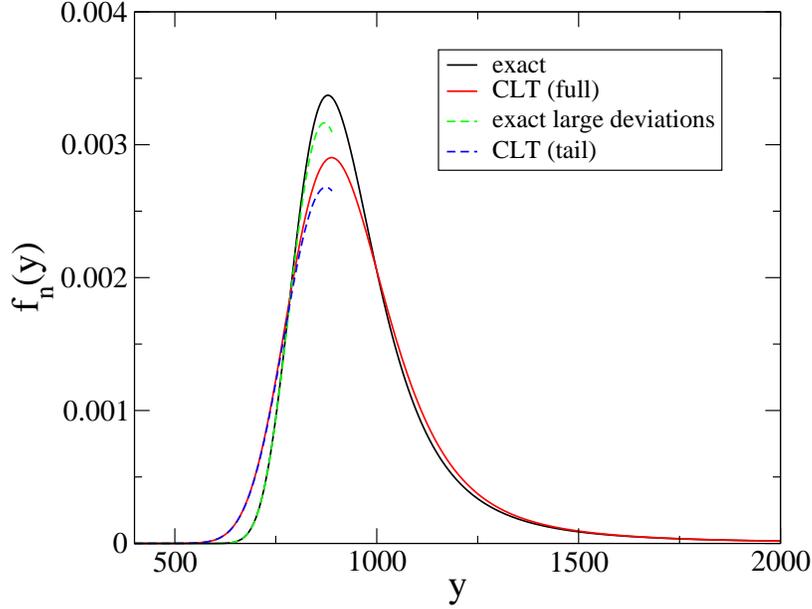}
\caption{\small
Density $f_{n}(y)$ of the sum $S_n$ of $n=500$ random variables with density (\ref{eq:proto32}) ($\theta=3/2$), against $y$.
We compare the exact density (\ref{eq:hermit}), the prediction (\ref{eq:CLT32}) of the generalised central limit theorem (\textsc{clt}), the exact large deviation estimate (\ref{eq:LD32}) and the estimate (\ref{eq:gclt32}) of this expression in the scaling regime $\rho\lesssim c_{1}=2$, which is also the left tail estimate of (\ref{eq:CLT32}).
(See the text for comments and the short summary below.)
}
\label{fig:final}
\end{center}
\end{figure}
\begin{figure}[htb]
\begin{center}
\includegraphics[angle=0,width=1\linewidth]{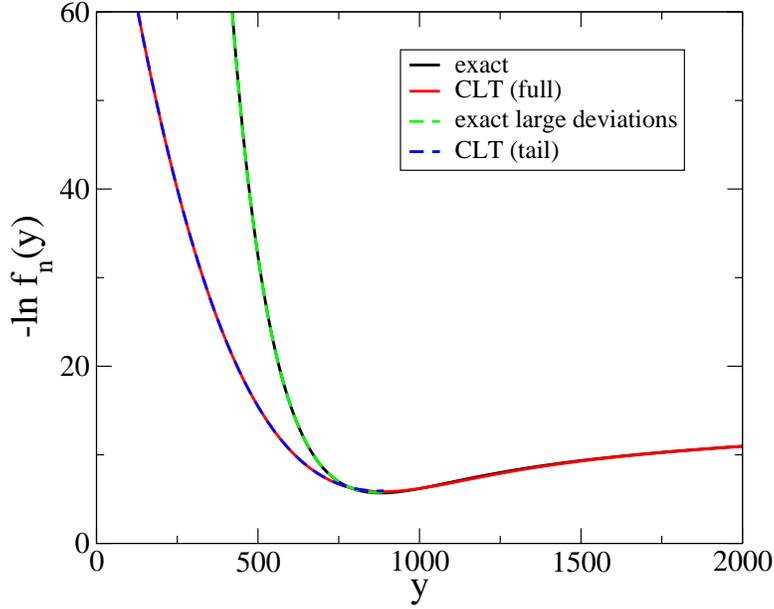}
\caption{\small
Same as figure~\ref{fig:final} in linear-log.
(See the text for comments.)
}
\label{fig:finallog}
\end{center}
\end{figure}

\subsubsection*{\CCB Central limit theorem.}
The generalised central limit theorem states that the bulk (i.e., for $\rho\approx 2$) of the distribution of the sum $S_n$ is given by (\ref{eq:gclt2}),
\beq\label{eq:CLT32}
f_{n}(y)\approx\frac{1}{n^{2/3}}L_{3/2,c}\left(\frac{y-2n}{n^{2/3}}\right),
\eeq
where the stable law $L_{3/2,c}(u)$ is explicitly known in terms of the Airy function and its derivative \cite{airy}.
With $c=2/\sqrt{\pi}$ it reads
\beq\label{eq:airy}
L_{3/2,c}(u)
=-\frac{1}{2}\exp\left(\frac{u^3}{96}\right)\left[\frac{u}{4}{\rm Ai}\left(\frac{u^2}{16}\right)
+{\rm Ai'}\left(\frac{u^2}{16}\right)\right],
\eeq
with Laplace transform
\be
\hat L_{3/2,c}(s)=\e^{\frac{8}{3}s^{3/2}}.
\ee

\subsubsection*{\CCB Left tail of (\ref{eq:CLT32}).}
For $u=(y-2n)/n^{2/3}$ large negative, 
 the behaviour of $L_{3/2,c}(u)$ is given by (\ref{eq:Lev32as}), thus, in this regime, (\ref{eq:CLT32}) yields
\beq\label{eq:gclt32}
f_{n}(y)\approx\frac{\sqrt{2-y/n}}{4\sqrt{n\pi}}\e^{-n(2-y/n)^3/48}.
\eeq
Comparing (\ref{eq:gclt32}) to the general expression (\ref{eq:largedev02}) yields the large deviation function $I(\rho)$ in the scaling regime $\rho\lesssim 2$,
\beq\label{eq:Irhobulk}
I(\rho)\approx\frac{1}{48}(2-\rho)^3.
\eeq
This expression, which is universal, is a particular form of (\ref{eq:largedev+}), with $\theta=3/2$ and $c_1=2$.

\subsubsection*{\CCB The large deviation function.}
Following the scheme given in section \ref{sec:leftLD} for the determination of the full large deviation function yields the saddle-point equation (\ref{eq:saddleGen}) 
\beq\label{eq:kprime32}
-K'( s_{\rho})=\frac{2}{1+2\sqrt{ s_{\rho}}}=\rho,
\eeq
hence
\beq\label{eq:ss32}
\sqrt{ s_{\rho}}=\frac{1}{\rho}-\frac{1}{2},
\eeq
confirming that the saddle point only exists for $\rho< c_{1}=2$.
For $\rho= c_{1}$, the saddle-point value $ s_{\rho}$ vanishes.
We thus find the expression of the large deviation function (as defined in (\ref{eq:largedev01})), which reads
\beq\label{eq:largedev}
-I(\rho)\equiv s_{\rho}\rho+K( s_{\rho})=\frac{\rho}{4}-\frac{1}{\rho}+\ln\frac{2}{\rho},\qquad (\rho<2),
\eeq
and $K''( s_{\rho})=\rho^3/(2-\rho)$.
Using (\ref{eq:largedev02}), we finally obtain
\beqa\label{eq:LD32}
f_{n}(y)
&\approx&
\frac{\sqrt{2-\rho}}{\sqrt{2\pi n\rho^3}}
\exp\left[n\left(\frac{\rho}{4}-\frac{1}{\rho}+\ln\frac{2}{\rho}\right)\right]
\\
&=&\frac{n(2n)^n\e^{-n^2/y+y/4}\sqrt{2-y/n}}{\sqrt{2\pi}y^{n+3/2}}.
\label{eq:largedev3+}
\eeqa
Two remarks are in order.
Firstly, for $\rho\to c_{1}$, i.e., $ s_{\rho}\to 0$, $K''( s_{\rho})\to\infty$.
The reason is that, according to (\ref{eq:Kpp}), $K''(0)=\var X$, which is infinite in the present case.
Hence one does not expect good accuracy of this prediction when approaching $ c_{1}$.
Secondly, the expansion of (\ref{eq:largedev}) for $\rho\lesssim 2$ yields (\ref{eq:Irhobulk}) as it should.
 In this regime the large deviation estimate (\ref{eq:LD32}) takes the universal form (\ref{eq:gclt32}).

\subsubsection*{\CCB Right tail of $f_n(y)$.}
When the difference $\Delta=y-2n$ is positive and extensive, the distribution of $S_n$ is given by (\ref{eq:deep}), 
with $c=2/\sqrt{\pi}$, that is
\beq\label{eq:deep32}
f_{n}(y)\approx\frac{2n}{\sqrt{\pi}(y-2n)^{5/2}}.
\eeq

\subsubsection*{\CCB Remark: asymptotics of $f_n(y)$ in the tails.}
The results (\ref{eq:LD32}) (left tail) and (\ref{eq:deep32}) (right tail) can also be obtained by a direct asymptotic analysis of the exact expression (\ref{eq:hermit}), as we now show.
In (\ref{eq:hermit}) the argument of the Hermite polynomial,
\be
z=\frac{y+2n}{2\sqrt{y}},
\ee
defines a function $z(y)$ which is minimum at $y=2n$, where $z=\sqrt{2n}$.
For $y$ smaller or greater than $2n$, $z$ is always larger than $\sqrt{2n}$.%
\footnote{In the language of a quantum harmonic oscillator, this means that the region explored in the variable $z$ when $y$ varies from zero to infinity is the forbidden region where the Hermite polynomials do not oscillate.}
We therefore need an asymptotic estimate of $H_n\left(z\right)$
for $z>\sqrt{2n}$.
This is obtained by a saddle-point analysis of the generating function of Hermite polynomials yielding (see \ref{app:Hermite})
\beq\label{eq:Hn-asymp}
\fl H_n(z)\approx
 \e^{(z^2-zV-n)/2}(z+V)^n\sqrt{(1+z/V)/2},\quad V=\sqrt{z^2-2n}.
\eeq
Using this estimate in (\ref{eq:hermit}),
then setting $y=n\rho$ with $\rho<2$, and expanding for $n\to\infty$
yields (\ref{eq:LD32}).
Likewise setting $y=n\rho$ with $\rho>2$, then expanding for $n\to\infty$,
yields (\ref{eq:deep32}).

\subsubsection*{\CCB Numerical comparisons of exact predictions and asymptotic estimates.}
In figures \ref{fig:final} and \ref{fig:finallog} we compare the analytical prediction (\ref{eq:hermit}) for the 
distribution of the sum of $n=500$ random variables with density (\ref{eq:proto32}) and tail index $\theta=3/2$, with 
\begin{description}
\item [$\star$] the prediction (\ref{eq:CLT32}) of the generalised central limit theorem,
\item [$\star$] the full large deviation estimate (\ref{eq:LD32}),
\item [$\star$] and the estimate (\ref{eq:gclt32}) of the latter in the scaling regime \textcolor{black}{$\rho\lesssim c_{1}=2$}; equation (\ref{eq:gclt32}) is equivalently the estimate for the left tail of the scaling form (\ref{eq:CLT32}).

\end{description}
These figures illustrate the following facts:
\begin{enumerate}
\item 
The right tail of the exact expression (\ref{eq:hermit}) is in excellent numerical agreement with the right tail of the scaling form (\ref{eq:CLT32}).
(See the comments below (\ref{eq:deep}).)
\item
The left tail of the exact expression (\ref{eq:hermit}) is in excellent numerical agreement with the large deviation estimate (\ref{eq:LD32}).
\item 
The left tail of the central limit expression (\ref{eq:CLT32}) is in excellent numerical agreement with its estimate (\ref{eq:gclt32}).

\end{enumerate}

\begin{figure}[htb]
\begin{center}
\includegraphics[angle=0,width=1\linewidth]{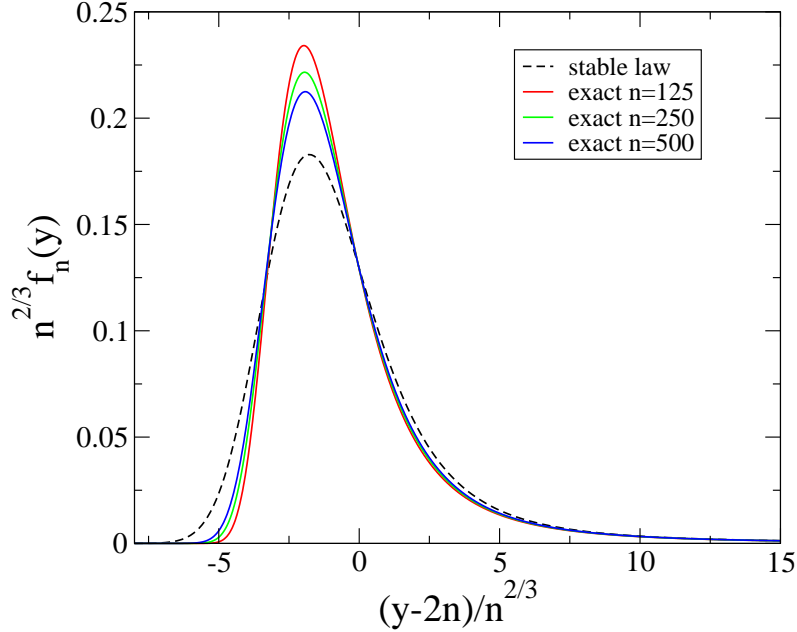}
\caption{\small
Comparison between the stable law (\ref{eq:airy}) and the exact result (\ref{eq:hermit}) centered and scaled for $n=125,250,500$.
}
\label{fig:convergNew}
\end{center}
\end{figure}

Figure~\ref{fig:convergNew} depicts the centered and scaled exact result (\ref{eq:hermit}) for $n=125,250,500$, together with the stable law (\ref{eq:airy}), illustrating the slow convergence of the former to the latter.

\subsubsection*{\CCB A short summary.}

The main equations obtained in this section and in section \ref{sec:power-law} can be summarised as follows,
\be
\left.\begin{array}{ccccccc}
\textsc{exact} && (\ref{eq:hermit}) &&-&&-
\\
\textsc{clt (full)}&& (\ref{eq:CLT32})&&(\ref{eq:gclt2})&& (\ref{eq:gclt1})
\\
\textsc{ld (full)}&& (\ref{eq:LD32})&&-&&-
\\ 
\textsc{deep}&& (\ref{eq:deep32})&&(\ref{eq:deep})&&(\ref{eq:deep})
\\ 
\left\{
\begin{array}{lr}
 \textsc{clt (tail)}& \\
 \textsc{ld (scaling)} & 
\end{array}
\right.
&& (\ref{eq:gclt32})&&(\ref{eq:largedev02}),(\ref{eq:largedev+})&&(\ref{eq:largedev02}),(\ref{eq:CLTback})
\end{array}\right.
\ee 
These equations are identified by short names or acronyms in the left column (\textsc{clt}: central limit theorem, \textsc{ld}: large deviation, \textsc{deep}: deep in the condensed phase).
The second column refers to results concerning the distribution (\ref{eq:proto32}), the third column refers to results concerning the generic case (\ref{eq:powerlaw}), with $1<\theta<2$, and the rightest column refers to the case $\theta>2$.

In the generic case (\ref{eq:powerlaw}) no exact expression for the distribution $f_n(y)$, as in (\ref{eq:hermit}), is known.
Neither is there in general an exact expression of the full large deviation estimate, as in (\ref{eq:LD32}).
The (generalised) central limit theorem reproduces correctly the behaviour of the left tail of $f_{n}(y)$ in the universal scaling region only, i.e., for $y$ close to $n c_{1}$.
For $\Delta$ negative and extensive, only the full large deviation estimate is faithful, which, as said above, is not explicitly known in general.
The right tail expression (\ref{eq:deep}) is valid for any $\theta>1$.

\subsubsection*{\CCB Comparison with the case of a discrete distribution.}

Finally, to complete this study, figure~\ref{fig:Z-H} depicts 
a comparison between the exact density $f_{n}(y)$ (\ref{eq:hermit}) and its discrete counterpart, the partition function $Z_{L,N}$ of the zero range process with hopping rate (\ref{eq:ukproto}), where $b=5/2$.
The partition function is obtained recursively using (\ref{eq:recursion+}).
The curves are centered and scaled, in order to highlight the universality of the bulk in the continuum limit.
The parameter $r$ is the ratio of the tail parameters of the two functions, namely $c=2/\sqrt{\pi}$ for the first one and $c=(b-1)\Gamma(b)=9\sqrt{\pi}/8$ for the second one (see (\ref{eq:fkc}) or (\ref{eq:zas})).
The lower plot demonstrates the non universality of the large deviations in the left tail.

%
\section{Marginal conditional density and condensation} 
\label{sec:marginal}

We are now in position to compute the marginal conditional distribution (\ref{eq:marginal}), repeated here for convenience,
\beq\label{eq:marginal-}
f(x|y)=f_X(x)\frac{f_{n-1}(y-x)}{f_{n}(y)},
\eeq
where the density $f_X(x)$ is given by (\ref{eq:powerlaw}).
This conditional density is a function of $x$, while $y$ plays the role of a parameter.
We thus have to study separately $f(x|y)$ for the different regimes of $y/n=\rho$.
The study hereafter parallels that made in \cite{maj2}.

\subsubsection*{\CCB Subcritical regime ($\rho< c_{1}\Leftrightarrow\Delta<0$).}
We start again from (\ref{eq:fSny}).
Thus
\be
f_{n-1}(y-x)=\int_{C}\frac{{\rm d}s}{2\pi{\rm i}}\e^{sy+n K(s)-sx-K(s)}.
\ee
Let us assume that $x$ is of order 1.
So, at the saddle point, for $n$ large, we have, within exponential accuracy (see section \ref{sec:leftLD}),
\beq\label{eq:fnm1}
f_{n-1}(y-x)\sim\e^{-nI(\rho)- s_{\rho} x-K( s_{\rho})},
\eeq
where $ s_{\rho}$ satisfies the equation $-K'( s_{\rho})=\rho$.
This yields, for any $\theta>1$, the handy expression
\beq\label{eq:marginal-sub}
f(x|y)\comport{\approx}{n\to\infty}{}\frac{f_X(x)\e^{- s_{\rho} x}}{\hat f_X( s_{\rho})},
\eeq
which is well normalised and has its first moment equal to $\rho$.
Its physical interpretation is appealing: there is `compression' of the $X_i$, since each one of them bears a part of the negative difference $\Delta$.
This accounts for the \textit{fluid phase}.
When $x$ becomes large (\ref{eq:fnm1}) and (\ref{eq:marginal-sub}) are no longer correct.
It is necessary to use the large deviation estimate (\ref{eq:largedev02}) in order to obtain an accurate expression of the marginal density (\ref{eq:marginal-}). 

This study can be illustrated on the example of $f_X(x)$ given by (\ref{eq:proto32}) ($\theta=3/2$).
Equation (\ref{eq:marginal-sub}) yields (using the accurate expression (\ref{eq:ss32}) for $ s_{\rho}$)
\beq\label{eq:marginal32}
f(x|y)\comport{\approx}{n\to\infty}{}\frac{\rho}{\sqrt{\pi} x^{5/2}}\exp\left(-\left[\frac{(2-\rho)x+2\rho}{2\rho \sqrt{x}}\right]^2\right).
\eeq
This expression is in excellent numerical agreement with the exact prediction for $f(x|y)$ derived from (\ref{eq:hermit}) if $x$ is of order 1, as soon as $n$ is large enough.
In contrast, the estimate obtained for $f(x|y)$ using the scaling estimate (\ref{eq:sstar}) for $ s_{\rho}$ compares well to the true distribution only when $\rho$ is not too far away from $ c_{1}$.
Finally, if $x$ is no longer of order 1, the large deviation estimate (\ref{eq:largedev3+}) inserted into (\ref{eq:marginal-}) provides an accurate estimate of the marginal distribution $f(x|y)$.
Starting from this very expression, setting $y=n\rho$ and letting $n\to\infty$ restores (\ref{eq:marginal32}),
since $x$ becomes $\ll y$ in this limit.

\subsubsection*{\CCB Critical regime ($\rho= c_{1}\Leftrightarrow\Delta=0$).}
Note that if $\rho= c_{1}=2$, then $ s_{\rho}=0$ and both asymptotic estimates (\ref{eq:marginal-sub})
and (\ref{eq:marginal32}) reduce to $f_X(x)$.
These estimates are obtained in the limit $n\to\infty$ (in order for the saddle-point method to be valid).
Therefore the reduction of $f(x|y)$ to $f_X(x)$ only holds in this limit.
Otherwise there are finite-size corrections given by the expressions (\ref{eq:marginal+}) and (\ref{eq:marginal++}) below, where the estimate of $f_n(y)$ in the bulk is used.
For $1<\theta<2$,
\beq\label{eq:marginal+}
f(x|y)\approx f_X(x)\frac{L_{\theta,c}((c_1-x)/n^{1/\theta})}{L_{\theta,c}(0)},
\eeq
and for $\theta>2$,
\beq\label{eq:marginal++}
f(x|y)\approx f_X(x)\frac{G((c_1-x)/n^{1/2})}{G(0)}=f_X(x)\e^{-(x-c_1)^2/2nc_2}.
\eeq
Again, if $n\to\infty$, one recovers the fact that $f(x|y)\to f_X(x)$.
For $x$ of order $n$, one should use the large deviation estimate (\ref{eq:largedev02}) for $f_{n-1}(y-x)$ (e.g. (\ref{eq:LD32}) for $f_X(x)$ given by (\ref{eq:proto32}), with $\theta=3/2$). 

\subsubsection*{\CCB Supercritical regime ($\rho> c_{1}\Leftrightarrow\Delta>0$).}
In this regime $f_{n}(y)$ is always given by its right-tail estimate (\ref{eq:deep}) 
\beq\label{eq:queue}
f_{n}(y)\approx\frac{nc}{\Delta^{1+\theta}}.
\eeq
The discussion therefore only focusses on $f_{n-1}(y-x)$, where $x$ should be compared to $\Delta$, which is of order $n$.
Beyond the obvious regime where $x$ is of order unity, hence $f(x|y)\approx f_X(x)$, there are three other regimes to consider, corresponding respectively to the bulk, the right-tail and the large deviations of $f_{n-1}(y-x)$.
\begin{description}
\item[(a) Condensate.] 
If $x\approx\Delta$ (that is $\Delta-x\sim n^{1/\alpha}$),
the ratio of $f_X(x)\approx c/\Delta^{1+\theta}$ to $f_{n}(y)$ given by (\ref{eq:queue}) yields
one piece of $f(x|y)$ 
\be
\frac{f_X(x)}{f_{n}(y)}\approx \frac{c/\Delta^{1/\theta}}{nc/\Delta^{1/\theta}}=\frac{1}{n}.
\ee
The other piece, $f_{n-1}(y-x)$, is given by its bulk since $y-x\approx nc_1$.
Hence, if $1<\theta<2$,
\beq\label{eq:bulkL}
\left. f(x|y)\right|_{\rm cond}\approx\frac{1}{n}f_{n-1}(y-x)\approx\frac{1}{n}\frac{1}{n^{1/\theta}}L_{\theta,c}\left(\frac{\Delta-x}{n^{1/\theta}}\right),
\eeq
and, if $\theta>2$,
\beq\label{eq:bulkG}
\left. f(x|y)\right |_{\rm cond}\approx\frac{1}{n}f_{n-1}(y-x)\approx\frac{1}{n}\frac{1}{n^{1/2}}G\left(\frac{\Delta-x}{n^{1/2}}\right).
\eeq
These expressions describe the bulk of the fluctuating condensate which manifests itself by a hump shape of the marginal $f(x|y)$ for $x\approx\Delta$ on figure \ref{fig:supercrit}.
For any $\theta>1$ we have, from (\ref{eq:bulkL}) or (\ref{eq:bulkG}),
\beq\label{eq:normalis}
\fl\int_{x\in\mathrm{hump}}{\rm d}x\,\left. f(x|y)\right|_{\rm cond}\approx
\frac{1}{n},
\eeq
which demonstrates that the excess difference $\Delta$ is borne by only one summand.
(See also the discussion in section \ref{sec:unique}.)

\item[(b) Dip.] 
The range of values of $x$ such that $x\gg1$, $\Delta-x\gg1$, interpolates between the critical part of $f(x|y)$, for $x$ or order $1$, and the condensate, for $x$ close to $\Delta$.
It corresponds to the \textit{dip region} on figure \ref{fig:supercrit}.
In this region, $f_{n-1}(y-x)$ is given by its right tail (\ref{eq:deep}) or (\ref{eq:queue}).
So, for any $\theta>1$,
\beq\label{eq:dip}
\left. f(x|y)\right|_{\rm dip}
 \approx 
 c\left[\frac{\Delta}{x(\Delta-x)}\right]^{1+\theta}
 \approx 
 \frac{f_X(x)f_X(\Delta-x)}{f_X(\Delta)}.
\eeq
The interpretation of this result is that in the dip region
typical configurations where one summand takes the value $x$ 
are such that the remaining $\Delta-x$ excess difference
is borne by a single other summand.
The dip region is therefore dominated
by configurations where the excess difference is shared by {\it two} summands \cite{gl2005}. 

The weight of these configurations can be estimated as follows.
Let $\xi$ be some positive number less than $1/2$.
Then
\beq\label{eq:dipW}
\fl\int_{\xi\Delta}^{(1-\xi)\Delta} {\rm d}x\,\left. f(x|y)\right|_{\rm dip}
=
\int_{\xi\Delta}^{(1-\xi)\Delta} {\rm d}x\,\frac{f_X(x)f_X(\Delta-x)}{f_X(\Delta)}
 \sim \Delta^{-\theta}\sim n^{-\theta}.
\eeq
The relative weights of the dip and condensate regions is therefore of order $n^{-(\theta-1)}$,
i.e., the weight of events where the condensate is broken in two pieces of order $n$ is subleading with respect to events with a single big jump.
This will be restated in section \ref{sec:unique}.
The reduction factor $n^{-(\theta-1)}$ is the same as that met in the discussion at the end of section \ref{sec:power-law}.

An illustration of this phenomenon is given in figure \ref{fig:RW}.
The overwhelming contribution to the statistics of trajectories comes from those exhibiting a single big jump of order $n$, approximately equal to $\Delta$.
Some rare trajectories, as the green one, exhibit two big jumps instead of a single one, both of order $n$.
These trajectories contribute to (\ref{eq:dip}).

\item[(c) Large deviations.] 
Finally, if $x>\Delta$, one should use the large deviation estimate for $f_{n-1}(y-x)$ (e.g. (\ref{eq:LD32}) for $f_X(x)$ given by (\ref{eq:proto32})). 
\end{description}

In summary, the contribution of the condensate to the total weight is equal to $1/n$.
The contribution of the dip region is subleading by a power-law factor.
The contribution of the large deviations is exponentially subleading.
The main contribution comes from the region where $x$ is of order unity where $f(x|y)\approx f_X(x)$.

\subsubsection*{\CCB Quantitative comparison.}
\begin{figure}[htb]
\begin{center}
\includegraphics[angle=0,width=1\linewidth]{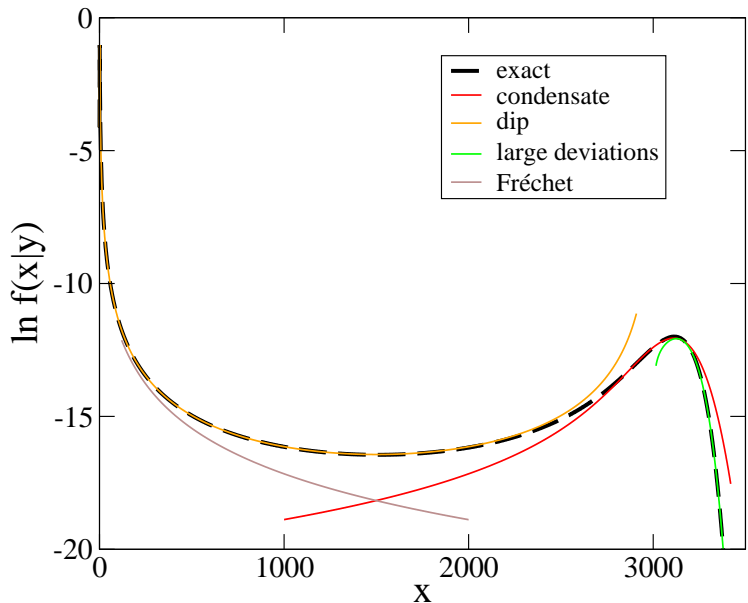}
\caption{\small
Linear-log plot of the marginal distribution $f(x|y)$ in the supercritical regime for $f_X(x)$ given by (\ref{eq:proto32}) ($\theta=3/2$).
Here $n=500$, $\rho=y/n=4 c_{1}=8$, hence $\Delta\equiv y-n c_{1}=3000$.
The exact marginal distribution $f(x| y)$ is reproduced by the union of three pieces, respectively the condensate, the dip and the large deviation contributions.
The curve named Fr\'echet represents $f^{(2)}(x|y)/n$ as defined in (\ref{eq:frech}) (see section \ref{sec:extremes}).
}
\label{fig:supercrit}
\end{center}
\end{figure}

Figure~\ref{fig:supercrit} summarises this study.
It depicts the marginal distribution $f(x|y)$, with $f_X(x)$ given by (\ref{eq:proto32}), for $n=500$, $y=4 nc_1$, $\Delta=y-nc_1=3000$ ($c_1=2$).
The curves named condensate, dip and large deviations correspond respectively to the cases (a), (b) and (c) above.
The curve named Fr\'echet represents $f^{(2)}(x)/n$ as defined in (\ref{eq:frech}) and
will be commented on in section \ref{sec:extremes}.

%
\section{Unicity of the condensate }
\label{sec:unique}

The analysis of the marginal distribution $f(x|y)$ made in section \ref{sec:marginal} showed that
the distribution $f(x|y)$ has a hump shape for $x\approx\Delta$,
the weight of which is equal to $1/n$ according to (\ref{eq:normalis}).
This means that the largest summand is the only one to `bear' the excess difference $\Delta$ and therefore that 
asymptotically the condensate is unique.

However, as discussed below (\ref{eq:dip}), 
there exist configurations where the excess difference is shared by two summands (i.e., with now a leader and a subleader instead of a unique condensate) 
and whose weight is subleading by a factor of order $n^{-(\theta-1)}$ with respect to configurations with a single big jump.
Such configurations are those which dominate in the dip region.

We present hereafter another argument in favour of the unicity of the condensate which is independent of that recalled above, even if it is akin to it.
The aim is to show that the event with a unique $X_i$ bearing all the excess difference $\Delta$ is much more likely than the event corresponding to two summands $X_i$ sharing it.
This issue has been previously discussed in \cite{evans2} for discrete variables, in the context of the statics of the zero-range process. 
Uniqueness of the condensate has also been established rigorously in the discrete and continuous cases in \cite{armendariz2009} and \cite{armendariz2011} (see also \cite{gross}). 
 
The probability associated to the event where $X_1$ bears the excess difference is
\beq\label{eq:1condensa}
\prob(\Delta<X_1<\Delta+{\rm d}x|S_n=y)= f(\Delta |y){\rm d}x,
\eeq
with
\beq\label{eq:1condensat}
f(\Delta |y)= f_X(\Delta)\frac{f_{n-1}(y-\Delta)}{f_{n}(y)},
\eeq
and where $y-\Delta=nc_1$.
This probability has to be multiplied by a factor $n$ since any of the $X_i$ can be chosen to bear the excess difference.

The probability corresponding to the event where $X_1$ and $X_2$ are both large and share the excess difference $\Delta$ reads
\beq\label{eq:2condens}
\fl\prob(\Delta<X_1+X_2<\Delta+{\rm d}x|S_n=y)=\left( \int_{\xi\Delta}^{(1-\xi)\Delta}{\rm d}x' f\left(x',\Delta-x'|y\right)\right){\rm d}x,
\eeq
with
\beq\label{eq:2condensa}
f\left(x,\Delta-x|y\right)=f_X(x)f_X(\Delta-x)\frac{f_{n-2}(y-\Delta)}{f_{n}(y)},
\eeq
and where $\xi$ is some positive number less than $1/2$, as in (\ref{eq:dipW}).
The probability (\ref{eq:2condens}) has to be multiplied by the binomial coefficient $\binom{n}{2}$ which counts the possible choices of two $X_i$ amongst $n$.
The ratio $f_{n-1}(nc_1)/f_{n-2}(nc_1)$ is asymptotically equal to one, so remains to estimate
\beq\label{eq:1cond}
nf_X(\Delta)\sim n\Delta^{-1-\theta}\sim n^{-(\theta-1)-1},
\eeq
and
\beq\label{eq:2cond}
\binom{n}{2}\int_{\xi\Delta}^{(1-\xi)\Delta} {\rm d}xf_X(x)f_X(\Delta-x)
 \sim n^2\Delta^{-1-2\theta}\sim n^{-2(\theta-1)-1}.
\eeq
The ratio of these two estimates scales as $n^{\theta-1}\gg1$ as soon as $\theta>1$, which is precisely the condition for the existence of a condensate. 
This result can be generalised to the case of $p$ variables sharing the excess difference $\Delta$.
We now have to estimate
\be
\fl \binom{n}{p}\int{\rm d}x_1\dots{\rm d}x_p\,f(x_1)\dots f(x_p)\delta\big(\sum_{i}x_i-\Delta\big)
\sim n^p\frac{\Delta^{p-1}}{\Delta^{p(1+\theta)}}\sim n^{-p(\theta-1)-1}.
\ee
Thus the ratio of (\ref{eq:1cond}) to the latter yields $n^{(p-1)(\theta-1)}$.

\subsubsection*{\CCB Remarks.}
\begin{enumerate}
\item The factor $n^{\theta-1}$ is precisely that found at the end of section \ref {sec:power-law} by a different line of reasoning.
\item Performing the integral in (\ref{eq:2cond}) from $0$ to $\Delta$ would yield $f_2(\Delta)$ which scales as $\Delta^{-1-\theta}$ instead of $\Delta^{-1-2\theta}$ as in (\ref{eq:2cond}).
Multiplied by $\binom{n}{2}$ this yields $n^{1-\theta}$, which dominates (\ref{eq:1cond}) by a factor $n$, as it should.
\item As a last remark, let us note that 
the ratio of (\ref{eq:1condensat}) to (\ref{eq:2condensa}) gives
\be
\frac{f(x,\Delta-x|y)}{f(\Delta|y)}\approx 
 \frac{f_X(x)f_X(\Delta-x)}{f_X(\Delta)},
\ee
which is the expression (\ref{eq:dip})
for $\left. f(x|y)\right|_{\rm dip}$.

\end{enumerate}

\section{Largest summands} 
\label{sec:extremes}
Investigating the statistics of extremes for the problem at hand is a natural question since the condensate is the largest summand.
A number of works have been devoted to this question \cite{jeon,ferrari, armendariz2009,maj3,janson}.
The discussion hereafter concerns the case where $f_X(x)$ has a power-law tail (\ref{eq:powerlaw}).

In \cite{jeon} the greatest summand is proven to scale as $n$ in the supercritical regime, as $n^{1/\theta}$ in the critical regime and as $\ln n$ in the subcritical regime.
In \cite{ferrari} it is shown that if the largest summand is removed, the measure on the remaining summands converges to the product measure with density $\rho=c_1$, when the number of summands $n$ is fixed and the value of the sum $y$ increases to infinity.
This means that the remaining background is critical, a feature which is apparent in figure \ref{fig:RW}, as already mentioned.

Let us denote the $k-$th largest summand by $X^{(k)}$ ($k=1,\dots,n$).
The densities of these ranked summands, denoted by $f^{(k)}(x|y)$, sum up to
\be
\sum_{k=1}^n f^{(k)}(x|y)=nf(x|y).
\ee
The distribution of the largest summand $X^{(1)}$ is investigated in \cite{maj3,armendariz2009,janson}. 
The result is that, if $y=n\rho$, $\rho>c_1$, $n\to\infty$, 
the rescaled variable
\be
Z_n=n^{-1/\alpha}(\Delta-X^{(1)})
\ee
converges to a stable law of index $\alpha$, with $\alpha$ defined in (\ref{eq:alfa}) 
(i.e., $\alpha=\theta$ if $\theta<2$ or $\alpha=2$ if $\theta>2$). 
This means that, asymptotically, the density of $X^{(1)}$ coincides, up to a factor $n$, with the estimates of the marginal density in the condensate region ($\Delta-x\sim n^{1/\alpha}$), that is with (\ref{eq:bulkL}) or (\ref{eq:bulkG}) according to the value of $\theta$, 
\be
f^{(1)}(x|y)\approx n\left.f(x|y)\right |_{\rm cond}, \qquad (\Delta-x\sim n^{1/\alpha}).
\ee
This result conforms with the intuition that, in the condensate region, 
the only contribution to the marginal $f(x|y)$ comes from the largest summand.

One can already guess from the statements made in \cite{jeon,ferrari} and recalled above that the distribution of the second largest summand, $X^{(2)}$, should be asymptotically Fr\'echet, and that the subsequent ones, 
$X^{(k)}$ 
$(k\ge 2)$, should be the order statistics of $n-1$ iid random variables $X_i$ with density $f_X(x)$ (i.e., before conditioning),
which can be summarised by saying that, in the supercritical regime, the dependency between the summands $X_i$ introduced by the conditioning goes asymptotically in the big jump $X^{(1)}$.
Reference \cite{janson} indeed states that the rescaled variables 
\be
W_n^{(k)}=n^{-1/\theta}X^{(k)}, \quad (k\ge 2),
\ee
have asymptotic densities
\be
f^{(k)}_W(w)=\frac{c}{w^{1+\theta}}\e^{-c/(\theta w^\theta)}\frac{\big[c/(\theta w^\theta)\big]^{k-2}}{(k-2)!},
\ee
independently of the value of $y$.
Hence, for $k\ge 2$,
\be
f^{(k)}(x|y)\approx\frac{1}{n^{1/\theta}}f^{(k)}_W\Big(\frac{x}{n^{1/\theta}}\Big)=
\frac{nc}{x^{1+\theta}}
\e^{-nc/(\theta x^\theta)}\frac{\big[nc/(\theta x^\theta)\big]^{k-2}}{(k-2)!}.
\ee
For instance the curve named Fr\'echet in figure \ref{fig:supercrit} represents 
\beq\label{eq:frech}
\frac{1}{n}f^{(2)}(x|y)\approx
\frac{2}{\sqrt{\pi}x^{5/2}}
\e^{-4n/(3\sqrt{\pi}x^{3/2})}.
\eeq

Since $X^{(1)}$ typically scales as $n$, while $X^{(2)}, X^{(3)},\dots$ typically scale as $n^{1/\theta}$,
the condensate is increasingly separated from the background as $n$ increases, leaving space to the dip region ($x\gg1$, $\Delta-x\gg1$).
We know from the analysis made in section \ref{sec:marginal} (see discussion following (\ref{eq:dip})) that 
this region is dominated
by configurations where the excess difference is shared by two summands, namely $X^{(1)}$ and $X^{(2)}$, so 
\beq\label{eq:f12dip}
 f^{(1)}(x|y)+f^{(2)}(x|y) \approx n\left.f(x|y)\right |_{\rm dip}, \quad (x\gg1, \Delta-x\gg1),
\eeq
and that the contributions of these events to $nf(x|y)$ are of order $n^{-(\theta-1)}$.
To the right of $\Delta/2$ the predominant contribution to the sum on the right side of (\ref{eq:f12dip}) comes from $ f^{(1)}(x|y)$, to the left it comes from
$ f^{(2)}(x|y)$.
In this respect it is worth noting that, right in the middle of the dip, i.e., for $x=\Delta/2$, the following relations hold, if $1<\theta<2$,\footnote{The crossing of
$\left. f(x|y)\right|_{\rm cond}$ and $f^{(2)}(x|y)/n$ at $x=\Delta/2$ is visible on figure \ref{fig:supercrit}.}

\bea
\left. f(x|y)\right|_{\rm dip}&\approx& 4^{1+\theta}\frac{c}{\Delta^{1+\theta}} ,
\\
\left. f(x|y)\right|_{\rm cond}&\approx&
\frac{1}{n}f^{(1)}(x|y)\
\approx
\frac{1}{n}f^{(2)}(x|y)
\\
&\approx& f_X(x)
\approx
2^{1+\theta}\frac{c}{\Delta^{1+\theta}},
\eea
where $\left. f(x|y)\right|_{\rm cond}$ is continued outside its region of validity ($\Delta-x\sim n^{1/\theta}$).
The ratio between the two quantities on the left side of the equations is therefore a universal number, only depending on the tail exponent $\theta$.
Up to adding a tail correction to $f^{(1)}(x|y)$ the same results are equally valid for $\theta>2$.

\subsubsection*{\CCB Remark.}
The random variable $Z_n$ is scaled by $n^{1/\alpha}$, where $\alpha$ is defined in (\ref{eq:alfa}), while the random variables $W_n^{(k)}$ are scaled by $n^{1/\theta}$.
In the first case $\alpha$ saturates at $\alpha=2$, in the second case $\theta$ can take any value.\footnote{Compare to theorem 19.34 in \cite{janson} where the distinction between these two exponents is not made.}
\section{Discussion}

In this work we have revisited the statistics of iid random variables with a power-law distribution (\ref{eq:powerlaw}) conditioned by the value of their sum.
For large values of the latter, a condensation transition occurs where the largest summand accommodates the excess difference between the value of the sum and its mean.
This simple scenario of condensation underlies a number of studies in statistical physics, usually formulated in terms of discrete random variables such as, e.g., in random allocation and urn models, or condensing zero-range processes at stationarity.
The present study extends easily to other subexponential distributions of the summands.

Much of the effort here has been devoted to presenting the subject in simple terms,
reproducing known results (especially from \cite{maj2} and \cite{gl2005}) and adding some new ones.
In particular the comparison between asymptotic estimates and their finite-size counterparts demonstrates the role of the contributions of the dip and large deviation regimes.
The contribution of the dip region is of crucial importance for the analysis of the stationary dynamics of the condensate \cite{gl2005}.
The conclusions given in \cite{gl2005} have been confirmed by rigorous mathematical studies \cite{meta1,meta2,meta3,meta4}.

To close, let us mention several related topics or generalisations of interest in \cite{sza,marsili,bertin,corberi}.

\ack
It is a pleasure to thank J M Luck for many interesting discussions, S N Majumdar for discussions on ref. \cite{maj2} and an anonymous referee for suggestions to improve the text.

\begin{appendix}
\section{Discrete formalism}
\label{app:discrete}

All the questions investigated so far with continuous random variables have a transcription in the language of discrete random variables.
The resulting framework is that used in the description of equilibrium urn models in statistical mechanics or in the analysis of the stationary state of zero range processes.
We successively review these three facets of the subject.
Table~\ref{tab:1} summarises the correspondences between the discrete and continuum formalisms.

\subsection{Discrete random variables conditioned by the value of their sum}
\label{app:discreterv}
Let $N_1,N_2,\dots,N_L$ be iid positive discrete random variables with distribution
\beq\label{eq:pik}
\pi_k=\prob(N_1=k),
\eeq
and average
\beq\label{eq:rhoc}
\langle N_1\rangle=\sum_{k\ge0}k\pi_k=\rho_c.
\eeq
The joint distribution of these random variables reads
\beq\label{eq:joint}
\prob(N_{1}=n_1,\dots,N_{L}=n_L)=
\pi_{n_1}\dots \pi_{n_L}.
\eeq
Assume now that their sum, denoted by $S_L$, is conditioned to be equal to $N$.
Then the joint distribution of $\{N_1,N_2,\dots,N_L\}$ and $S_L$ is
\beq\label{eq:joint+}
\fl \prob(N_{1}=n_1,\dots,N_{L}=n_L, S_L=N)=
\pi_{n_1}\dots \pi_{n_L}
\delta\Big(\sum_{i} n_i,N\Big).
\eeq
Summing this expression on $n_1,\dots,n_L$ yields the distribution of $S_L$, or partition function $\tilde Z_{L,N}$, 
\beqa\label{eq:zustand}
\tilde Z_{L,N}&\equiv&\prob(S_L=N)
\nonumber\\
&=&\sum_{\{n_i\}}\pi_{n_1}\dots \pi_{n_L} \delta\Big(\sum_{i} n_i,N\Big).
\eeqa
The conditional joint distribution of $N_1,N_2,\dots,N_L$, given $S_L$, is the ratio of (\ref{eq:joint+}) to (\ref{eq:zustand}), that is
\beq\label{eq:prob}
 \fl \prob(N_1=n_1,\dots,N_{L}=n_L|S_L=N)
=\frac{1}
{\tilde Z_{L,N}}\pi_{n_1}\dots \pi_{n_L} \delta\left(\sum_i n_i,N\right),
\eeq
from which the marginal conditional distribution of one of the $N_i$ (taken conventionally to be $N_1$), denoted by $f_k$, ensues by summation
\beq\label{eq:fkRV}
f_k=\prob(N_1=k|S_L=N)=
\pi_{k}
\frac{\tilde Z_{L-1,N-k}}{\tilde Z_{L,N}}.
\eeq
The conditional average is thus
\beq\label{eq:rhoconst}
\langle N_1|S_L=N\rangle=\sum_{k\ge0}k f_k=\frac{N}{L}=\rho,
\eeq
by definition of the density $\rho$.
Summing (\ref{eq:fkRV}) on $k$ leads to a recursion relation on the $\tilde Z_{L,N}$
\beq\label{eq:recursion}
\tilde Z_{L,N}=\sum_{k\ge0}\pi_k \tilde Z_{L-1,N-k}.
\eeq

\begin{figure}[htb]
\begin{center}
\includegraphics[angle=0,width=0.8\linewidth]{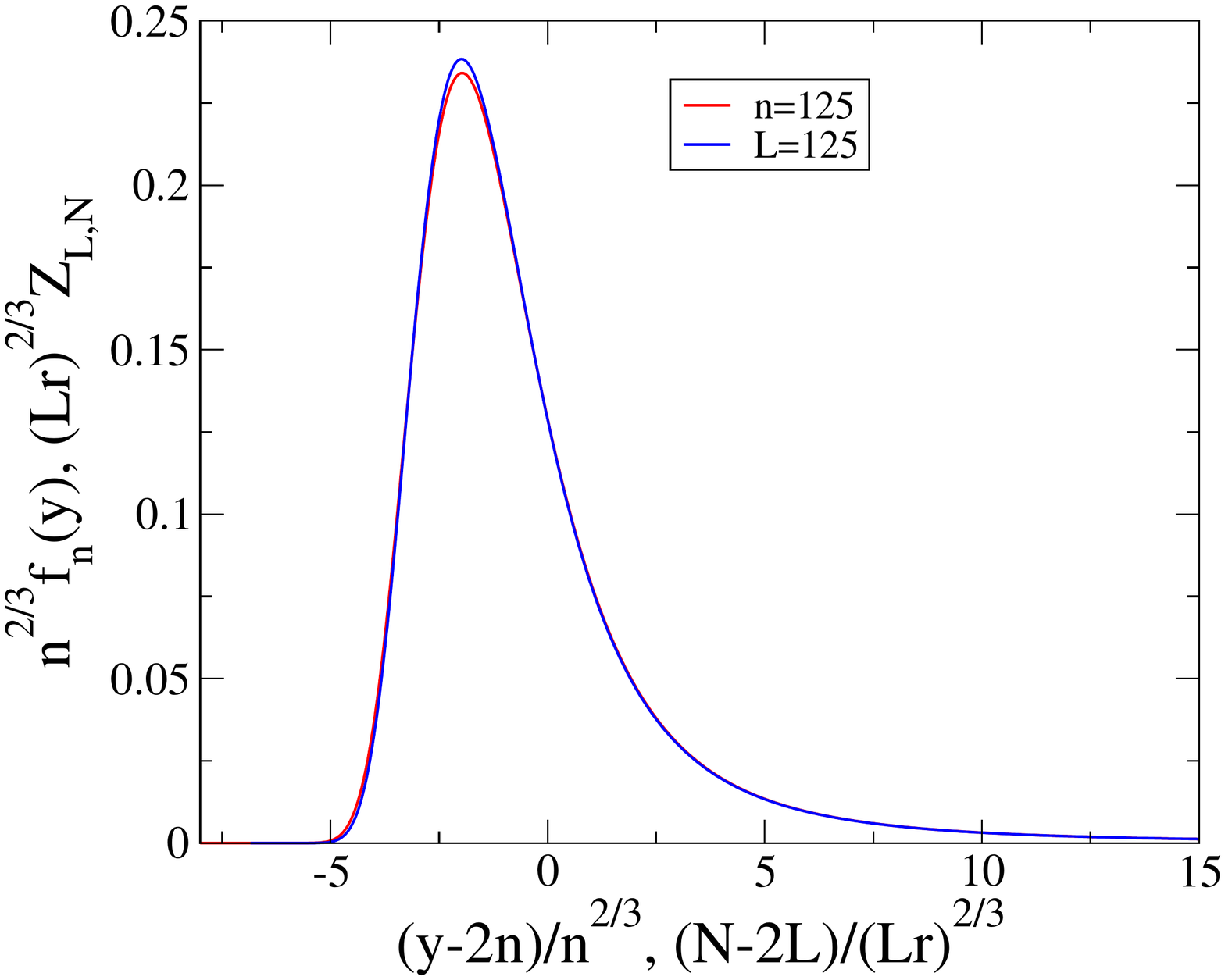}
\includegraphics[angle=0,width=0.8\linewidth]{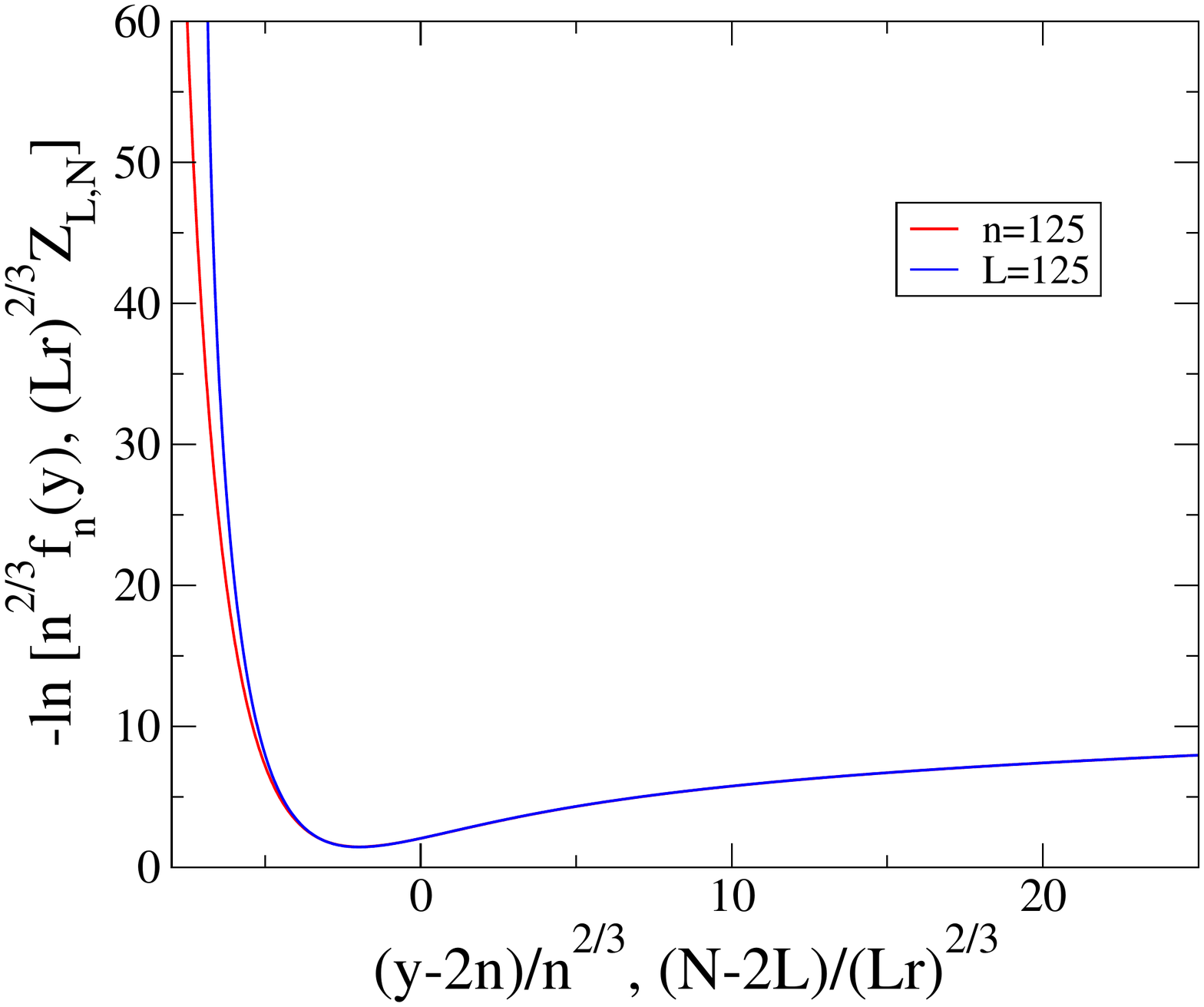}
\caption{\small
Comparison between the exact density $f_{n}(y)$ (\ref{eq:hermit}), for $n=125$, and the partition function $Z_{L,N}$ of the ZRP with hopping rate (\ref{eq:ukproto}) where $b=5/2$, for $L=125$, obtained recursively using (\ref{eq:recursion+}).
The curves are centered and scaled.
The parameter $r$ is the ratio of the tail parameters of the two functions (see text at the end of section \ref{sec:theta32}). 
The lower figure is a linear-log plot of the upper one.
}
\label{fig:Z-H}
\end{center}
\end{figure}

\begin{table}[ht]
\caption{Correspondences between the discrete formalism of \ref{app:discreterv} and the continuum formalism used in the bulk of the paper.}
\label{tab:1}
\begin{center}
\begin{tabular}{|c|c|}
\hline
Discrete r.v.&Continuous r.v.\\
\hline
$L$&$n$\\
$N_1,\dots,N_L$&$X_1,\dots,X_n$\\
$\prob(N_i=k)=\pi_k$&$f_X(x)$\\
$\langle N_1\rangle=\rho_c$&$\langle X\rangle=c_1$\\
$S_L=N_1+\cdots+N_L$&$S_n=X_1+\cdots+X_n$\\
$N$&$y$\\
$\rho=N/L$&$\rho=y/n$\\
$\prob(S_L=N)=\tilde Z_{L,N}$&$f_{n}(y)$\\
$f_k=\prob(N_1=k|S_L=N)$&$f(x|y)$\\
$\Pi(z)$&$\hat f_X(s)$\\
$\tilde Z_{L,N}\sim\e^{-L\mathcal{F}(\rho)}$&$ f_n(y)\sim\e^{-nI(\rho)}$\\
\hline
\end{tabular}
\end{center}
\end{table}

\subsubsection*{\CCB Thermodynamic limit.}
In the thermodynamic limit the large deviation function (or free energy) reads
\be
{\cal F}=-\lim_{L,N\to\infty}\frac{1}{L}\ln \tilde Z_{L,N},
\ee
i.e., with exponential accuracy,
\be
\tilde Z_{L,N}=\prob(S_L=N)\sim\e^{-L\,{\cal F}(\rho=N/L)}.
\ee
The large deviation function can be computed by the saddle-point method.
Casting the integral representation of the Kronecker function
\be
\delta(m,n)=\oint\frac{{\rm d}z}{2\pi{\rm i} z^{n+1}}\,z^m,
\ee
in (\ref{eq:zustand}) yields
\beq\label{eq:contour}
\tilde Z_{L,N}=\prob(S_L=N)=\oint\frac{{\rm d}z}{2\pi{\rm i} z^{N+1}}\,\Pi(z)^L,
\eeq
where $\Pi(z)$ is the generating function of the $\pi_k$ 
\be
\Pi(z)=\mean{z^{N_1}}=\sum_{k} z^k \pi_k.
\ee
The contour integral in~(\ref{eq:contour}) can be evaluated by the saddle-point method.
The saddle-point equation is
\be
\frac{z_{\rho}\, \Pi'(z_{\rho})}{\Pi(z_{\rho})}=\rho,
\ee
where the saddle-point value $z_{\rho}$ depends on the density $\rho$ through this equation.
The discussion of this equation is analogous to that given in the continuum formalism.

\subsection{Equilibrium urn models}

The framework described in the previous section is naturally realised by classical urn models, defined as follows.
Consider a finite connected graph, made of~$L$ sites (or urns),
on which $N$ particles are distributed.
The number of particles on site $i$ is the random variable $N_i$,
with $S_L=\sum_{i=1}^L N_i=N$.
A configuration of the system is defined by the values $\{n_1,\ldots ,n_L \}$,
taken by the random occupations $N_1,\dots,N_L$.
The energy of such a configuration 
is the sum of the individual energies at each site,
\be
E(\{n_i\})=\sum_{i=1}^{L}E(n_{i}).
\ee
The associated unnormalised Boltzmann weight attached to site $i$ is
\be
p_{n_{i}}=\e^{-\beta E(n_i)}.
\ee
The probability of the configuration $\{n_i\}$ is therefore 
given by the product form
\beq\label{eq:prob+++}
\fl\prob(N_{1}=n_1,\dots,N_{L}=n_L|S_L=N)=\frac{1}{ Z_{L,N}}
p_{n_{1}}\cdots p_{n_{L}}
\delta\Big(\sum_{i}n_{i},N\Big),
\eeq
where
\beq\label{eq:zustand+++}
Z_{L,N}=\sum_{\{n_i\}}\,p_{n_{1}}\cdots p_{n_{L}}\;
\delta\Big(\sum_{i}n_{i},N\Big),
\eeq
is the canonical partition function of this statistical mechanical system.
The single-site occupation probability is
\beq\label{eq:fkurn}
f_k=\prob(N_1=k|S_L=N)=
p_{k}
\frac{ Z_{L-1,N-k}}{ Z_{L,N}},
\eeq
and the partition function obeys the recursion relation 
\beq\label{eq:recursion+}
Z_{L,N}=\sum_{k\ge0}p_k Z_{L-1,N-k}.
\eeq

In order to make the link between the results of this section and those of \ref{app:discreterv} one normalises the $p_k$ as
\be
\pi_k=\frac{p_k}{\sum_{k}p_k},
\ee 
whenever the denominator is finite, thus recovering
the probabilities $\pi_k$ defined in (\ref{eq:pik}).
So doing,
(\ref{eq:zustand+++}) is proportional to (\ref{eq:zustand}) and
there is identity between (\ref{eq:prob+++}) and (\ref{eq:prob}), (\ref{eq:fkurn}) and (\ref{eq:fkRV})
and (\ref{eq:recursion+}) and (\ref{eq:recursion}).
For instance the `balls-in-boxes' model \cite{burda} has energy function
\be
E(n_i)=\ln(n_i+1),
\ee
yielding 
\be
p_k=\frac{1}{(1+k)^{\beta}},\qquad \pi_k=\frac{1}{\zeta(\beta)}\frac{1}{(1+k)^{\beta}},
\ee
where $\sum_{k}p_k=\zeta(\beta)$ is the Riemann zeta-function.
This model is the discrete counterpart of the case considered in the bulk of the paper where $f_X(x)$ has a power-law tail (\ref{eq:powerlaw}).
Here $\pi_k\sim k^{-\beta}$, with $\beta$ playing the role of $1+\theta$.

\subsection{Zero range process}
\label{app:zrp}

\subsubsection*{\CCB Definition.}
The zero range process can be seen as a dynamical extension of the class of static urn models discussed above.
We again consider a finite connected graph, made of $L$ sites.
At any time $t$ a configuration of the system is specified by the values taken by the occupation numbers 
$N_{i}(t)$, now functions of time.
The dynamics of the system consists in transferring
a particle from 
the departure site with label $d$, containing $N_{d}=k$ particles, to the
arrival site with label $a$ containing $N_{a}=\l$ particles.
By definition of a ZRP, the transfer rate is 
\be
W(d,a,k)=w_{d,a} u_k ,
\ee
where $u_k$ only depends on the occupation $N_d=k$ of the departure site and $w_{d,a}$ accounts for diffusion from site $d$ to site $a$.
To simplify, let us restrict the discussion to diffusion processes such that the stationary state is uniform.
The stationary probability of a configuration has the product form (\ref{eq:prob+++})
where the factor $p_{k}$ obeys the condition $p_k u_k=p_{k-1}$,
which gives the explicit form
\be
p_k=\frac{1}{u_1\dots u_k}.
\ee
The statics of this ZRP is therefore the same as that of the urn model sharing the same $p_k$.
Its partition function (\ref{eq:zustand+++}) obeys the recursion relation (\ref{eq:recursion+}) and 
the stationary single-site occupation probability is given by (\ref{eq:fkurn}).

Conversely, given an urn model, the corresponding ZRP has hopping rate $u_k=p_{k-1}/p_k$.
For the balls-in-boxes model \cite{burda} this yields \cite{camia,lux}
\be
u_k=\left ( 1+\frac{1}{k}\right)^\beta\approx 1+\frac{\beta}{k}.
\ee 

\subsubsection*{\CCB A prototypical condensing ZRP.}
\label{sec:ZRPproto}

The model with hopping rate 
\beq\label{eq:ukproto}
u_k=1+\frac{b}{k}
\eeq
is a well studied example of condensing ZRP.
The weights $p_k$ are given by
\be
p_k=\frac{\Gamma(b+1)\,k!}{\Gamma(k+b+1)}
=\int_0^1 {\rm d}u\,u^k\,b(1-u)^{b-1}\approx\frac{\Gamma(b+1)}{k^b},
\ee
with generating function
\be
P(z)=\sum_{k\ge0}z^kp_k=\int_0^1{\rm d}u\,\frac{b(1-u)^{b-1}}{1-zu}
=\null_2F_1(1,1;b+1;z),
\ee
where $_2F_1$ is the hypergeometric function.
This function has a branch cut at $z=z_c=1$, with a singular part of the 
form
\be
P_{\rm sg}(z)\approx A\,P(1)(1-z)^{b-1},\qquad A=\frad{(b-1)\pi}{\sin\pi b}
\ee
so that $P(z)$ is only differentiable $n\equiv \mathrm{Int}(b)-1$ many times at $z=z_c=1$:
\be
\fl P(z)\approx P(1)+(z-1)\,P'(1)+\cdots +\frac{(z-1)^n}{n!}P^{(n)}(1)+P_{\rm sg}(z),
\ee
with
\be
P(1)=\frad{b}{b-1},\quad
P'(1)=\frad{b}{(b-1)(b-2)},\dots
\ee

In the thermodynamic limit ($L\to\infty$ at fixed density $N/L=\rho$),
the system has a continuous phase transition at the critical density
\be
\rho_c=\frac{P'(1)}{P(1)}=\frac{1}{b-2},
\ee
whenever $b>2$. 
The critical density separates a fluid phase from a condensed phase.
In the fluid phase $(\rho<\rho_c)$,
the occupation probabilities $f_{k}$ fall off exponentially.
At the critical density $(\rho=\rho_c)$, they fall off as a power law:
\beq\label{eq:fkc}
f_{k}=\frac{p_k}{P(1)}\approx\frac{(b-1)\Gamma(b)}{k^b}.
\eeq
In the condensed phase $(\rho>\rho_c)$, for a large and finite system,
the particles form a uniform critical background
and a macroscopic condensate, consisting (on average) of
$\Delta$ excess particles with respect to the critical state, where
\be
\Delta=N-L\rho_c=L(\rho-\rho_c).
\ee
The condensate appears as a hump in the stationary distribution
$f_{k}$.
The expression of the partition function $Z_{L,N}$
{\it deep in the condensed phase}, i.e., for $\Delta=L(\rho-\rho_c)\gg1$
is \cite{gl2005}
\beq\label{eq:zas}
Z_{L,N}\approx(b-1)\Gamma(b)\frac{L}{\Delta^b}P(1)^L.
\eeq

\section{Asymptotics of Hermite polynomials}
\label{app:Hermite}

We want to demonstrate (\ref{eq:Hn-asymp}) which holds if $z>\sqrt{2n}$.
The generating function of the Hermite polynomials, defined as
\be
H_n(z)=(-)^n\e^{z^2}\frac{{\rm d}^n}{{\rm d}z^n}\e^{-z^2},
\ee
is
\be
\sum_{n\ge0}H_n(z)\frac{u^n}{n!}=\e^{2zu-u^2}.
\ee
Performing a saddle-point expansion of
\be
H_n(z)=n!\int\frac{{\rm d}u}{2\pi{\rm i} u}\e^{2zu-u^2-n\ln u},
\ee
one finds the saddle point
\be
u_c=\frac{z-V}{2},\qquad V=\sqrt{z^2-2n},
\ee
finally yielding
\be
H_n(z)\approx 
\e^{(z^2-zV-n)/2}(z+V)^n\sqrt{(1+z/V)/2}, \qquad (z>\sqrt{2n}),
\ee
which is (\ref{eq:Hn-asymp}).
\end{appendix}

\newpage

\end{document}